\PassOptionsToPackage{unicode}{hyperref}
\PassOptionsToPackage{hyphens}{url}
\PassOptionsToPackage{dvipsnames,svgnames,x11names}{xcolor}
\documentclass[
  11pt,
  a4paper,
]{article}

\usepackage{amsmath,amssymb}
\usepackage{iftex}
\ifPDFTeX
  \usepackage[T1]{fontenc}
  \usepackage[utf8]{inputenc}
  \usepackage{textcomp} 
\else 
  \usepackage{unicode-math}
  \defaultfontfeatures{Scale=MatchLowercase}
  \defaultfontfeatures[\rmfamily]{Ligatures=TeX,Scale=1}
\fi
\usepackage[]{libertinus}
\ifPDFTeX\else  
\fi
\IfFileExists{upquote.sty}{\usepackage{upquote}}{}
\IfFileExists{microtype.sty}{
  \usepackage[]{microtype}
  \UseMicrotypeSet[protrusion]{basicmath} 
}{}
\makeatletter
\@ifundefined{KOMAClassName}{
  \IfFileExists{parskip.sty}{%
    \usepackage{parskip}
  }{
    \setlength{\parindent}{0pt}
    \setlength{\parskip}{6pt plus 2pt minus 1pt}}
}{
  \KOMAoptions{parskip=half}}
\makeatother
\usepackage{xcolor}
\usepackage[a4paper,textheight=24cm,textwidth=15.5cm]{geometry}
\ifx\paragraph\undefined\else
  \let\oldparagraph\paragraph
  \renewcommand{\paragraph}[1]{\oldparagraph{#1}\mbox{}}
\fi
\ifx\subparagraph\undefined\else
  \let\oldsubparagraph\subparagraph
  \renewcommand{\subparagraph}[1]{\oldsubparagraph{#1}\mbox{}}
\fi

\usepackage{color}
\usepackage{fancyvrb}

\DefineVerbatimEnvironment{Highlighting}{Verbatim}{commandchars=\\\{\}}
\newenvironment{Shaded}{}{}

\newcommand{\AttributeTok}[1]{\textcolor[rgb]{0.84,0.23,0.29}{#1}}

\newcommand{\CommentTok}[1]{\textcolor[rgb]{0.42,0.45,0.49}{#1}}

\newcommand{\ConstantTok}[1]{\textcolor[rgb]{0.00,0.36,0.77}{#1}}
\newcommand{\ControlFlowTok}[1]{\textcolor[rgb]{0.84,0.23,0.29}{#1}}

\newcommand{\DecValTok}[1]{\textcolor[rgb]{0.00,0.36,0.77}{#1}}

\newcommand{\FloatTok}[1]{\textcolor[rgb]{0.00,0.36,0.77}{#1}}
\newcommand{\FunctionTok}[1]{\textcolor[rgb]{0.44,0.26,0.76}{#1}}

\newcommand{\NormalTok}[1]{\textcolor[rgb]{0.14,0.16,0.18}{#1}}

\newcommand{\OtherTok}[1]{\textcolor[rgb]{0.44,0.26,0.76}{#1}}

\newcommand{\SpecialCharTok}[1]{\textcolor[rgb]{0.00,0.36,0.77}{#1}}

\newcommand{\StringTok}[1]{\textcolor[rgb]{0.01,0.18,0.38}{#1}}

\providecommand{\tightlist}{%
  \setlength{\itemsep}{0pt}\setlength{\parskip}{0pt}}\usepackage{longtable,booktabs,array}
\usepackage{calc} 
\usepackage{etoolbox}
\makeatletter
\patchcmd\longtable{\par}{\if@noskipsec\mbox{}\fi\par}{}{}
\makeatother
\IfFileExists{footnotehyper.sty}{\usepackage{footnotehyper}}{\usepackage{footnote}}
\makesavenoteenv{longtable}
\usepackage{graphicx}
\makeatletter
\def\maxwidth{\ifdim\Gin@nat@width>\linewidth\linewidth\else\Gin@nat@width\fi}
\def\maxheight{\ifdim\Gin@nat@height>\textheight\textheight\else\Gin@nat@height\fi}
\makeatother
\setkeys{Gin}{width=\maxwidth,height=\maxheight,keepaspectratio}
\makeatletter
\def\fps@figure{htbp}
\makeatother
\newlength{\cslhangindent}
\newlength{\csllabelwidth}
\newlength{\cslentryspacingunit} 
\newenvironment{CSLReferences}[2] 
 {
  \setlength{\parindent}{0pt}
  \ifodd #1
  \let\oldpar\par
  \def\par{\hangindent=\cslhangindent\oldpar}
  \fi
  \setlength{\parskip}{#2\cslentryspacingunit}
 }%
 {}
\usepackage{calc}

\usepackage{tikz}
\usepackage{xcolor}
\usepackage{tabularx}
\usepackage{orcidlink}
\usepackage{stmaryrd}
\usepackage{xfrac}
\makeatletter
\@ifpackageloaded{caption}{}{\usepackage{caption}}
\AtBeginDocument{%
\ifdefined\contentsname
  \renewcommand*\contentsname{Table of contents}
\else
  \newcommand\contentsname{Table of contents}
\fi
\ifdefined\listfigurename
  \renewcommand*\listfigurename{List of Figures}
\else
  \newcommand\listfigurename{List of Figures}
\fi
\ifdefined\listtablename
  \renewcommand*\listtablename{List of Tables}
\else
  \newcommand\listtablename{List of Tables}
\fi
\ifdefined\figurename
  \renewcommand*\figurename{Figure}
\else
  \newcommand\figurename{Figure}
\fi
\ifdefined\tablename
  \renewcommand*\tablename{Table}
\else
  \newcommand\tablename{Table}
\fi
}
\@ifpackageloaded{float}{}{\usepackage{float}}
\floatstyle{ruled}
\@ifundefined{c@chapter}{\newfloat{codelisting}{h}{lop}}{\newfloat{codelisting}{h}{lop}[chapter]}
\floatname{codelisting}{Listing}

\makeatother
\makeatletter
\@ifpackageloaded{caption}{}{\usepackage{caption}}
\@ifpackageloaded{subcaption}{}{\usepackage{subcaption}}
\makeatother
\makeatletter
\@ifpackageloaded{tcolorbox}{}{\usepackage[skins,breakable]{tcolorbox}}
\makeatother
\makeatletter
\@ifundefined{shadecolor}{\definecolor{shadecolor}{rgb}{.97, .97, .97}}
\makeatother
\makeatletter
\@ifpackageloaded{algorithm}{}{\usepackage{algorithm}}
\makeatother
\makeatletter
\@ifpackageloaded{algpseudocode}{}{\usepackage{algpseudocode}}
\makeatother
\ifLuaTeX
  \usepackage{selnolig}  
\fi
\IfFileExists{bookmark.sty}{\usepackage{bookmark}}{\usepackage{hyperref}}
\IfFileExists{xurl.sty}{\usepackage{xurl}}{} 
\hypersetup{
  pdftitle={AdaptiveConformal: An R Package for Adaptive Conformal Inference},
  pdfauthor={Herbert Susmann; Antoine Chambaz; Julie Josse},
  pdfkeywords={Conformal inference, Adaptive conformal inference, time
series, R},
  colorlinks=true,
  linkcolor={blue},
  filecolor={Maroon},
  citecolor={Blue},
  urlcolor={Blue},
  pdfcreator={LaTeX via pandoc}}

\title{AdaptiveConformal: An \texttt{R} Package for Adaptive Conformal
Inference}
\usepackage{etoolbox}
\makeatletter
\providecommand{\subtitle}[1]{
  \apptocmd{\@title}{\par {\large #1 \par}}{}{}
}
\makeatother
\subtitle{AdaptiveConformal: An \texttt{R} Package for Adaptive
Conformal Inference}
\author{Herbert Susmann \and Antoine Chambaz \and Julie Josse}
\date{2023-11-30}

\begin{document}
\definecolor{computo-blue}{HTML}{034E79}

\begin{tikzpicture}[remember picture,overlay]
\fill[computo-blue]
  (current page.north west) -- (current page.north east) --
  ([yshift=-5cm]current page.north east|-current page.north east) --
  ([yshift=-5cm]current page.north west|-current page.north west) -- cycle;
\node[font=\sffamily\bfseries\color{white},anchor=west,
  xshift=4.25cm,yshift=-2.75cm] at (current page.north west)
  {\begin{minipage}{15cm}
    \fontsize{25}{30}\selectfont
    AdaptiveConformal: An \texttt{R} Package for Adaptive Conformal
    Inference
    \vspace{.5cm}
    \\
    \fontsize{15}{18}\selectfont
    AdaptiveConformal: An \texttt{R} Package for Adaptive Conformal
    Inference
  \end{minipage}};
\end{tikzpicture}

\vspace*{2.5cm}
\begin{center}
          Herbert
Susmann~\orcidlink{0000-0002-3540-8255}\footnote{Corresponding author: \href{mailto:herbert.susmann@dauphine.psl.eu}{herbert.susmann@dauphine.psl.eu}}\quad
             CEREMADE (UMR 7534), Université Paris-Dauphine PSL, Place
du Maréchal de Lattre de Tassigny, Paris, 75016, France\\
                 Antoine Chambaz~\orcidlink{0000-0002-5592-6471}\quad
             Université Paris Cité, CNRS, MAP5, F-75006 Paris, France\\
                 Julie Josse~\orcidlink{0000-0001-9547-891X}\quad
             Inria PreMeDICaL team, Idesp, Université de Montpellier\\
           
  \bigskip
  
  Date published: 2023-11-30 \quad Last modified: 2023-11-30
\end{center}
      
\bigskip
\begin{abstract}
Conformal Inference (CI) is a popular approach for generating finite
sample prediction intervals based on the output of any point prediction
method when data are exchangeable. Adaptive Conformal Inference (ACI)
algorithms extend CI to the case of sequentially observed data, such as
time series, and exhibit strong theoretical guarantees without having to
assume exchangeability of the observed data. The common thread that
unites algorithms in the ACI family is that they adaptively adjust the
width of the generated prediction intervals in response to the observed
data. We provide a detailed description of five ACI algorithms and their
theoretical guarantees, and test their performance in simulation
studies. We then present a case study of producing prediction intervals
for influenza incidence in the United States based on black-box point
forecasts. Implementations of all the algorithms are released as an
open-source \texttt{R} package, \texttt{AdaptiveConformal}, which also
includes tools for visualizing and summarizing conformal prediction
intervals.
\end{abstract}

\noindent%
\vfill
\ifdefined\Shaded\renewenvironment{Shaded}{\begin{tcolorbox}[boxrule=0pt, borderline west={3pt}{0pt}{shadecolor}, enhanced, sharp corners, frame hidden, breakable, interior hidden]}{\end{tcolorbox}}\fi

\floatname{algorithm}{Algorithm}

\floatname{algorithm}{Algorithm}

\floatname{algorithm}{Algorithm}

\floatname{algorithm}{Algorithm}

\floatname{algorithm}{Algorithm}

\renewcommand*\contentsname{Contents}
{
\hypersetup{linkcolor=}
\setcounter{tocdepth}{3}
\tableofcontents
}
\hypertarget{introduction}{%
\section{Introduction}\label{introduction}}

Conformal Inference (CI) is a family of methods for generating finite
sample prediction intervals around point predictions when data are
exchangeable (Vovk, Gammerman, and Shafer 2005; Shafer and Vovk 2008;
Angelopoulos and Bates 2023). The input point predictions can be derived
from any prediction method, making CI a powerful tool for augmenting
black-box prediction algorithms with prediction intervals. Classical CI
methods are able to yield marginally valid intervals with only the
assumption that the joint distribution of the data does not change based
on the order of the observations (that is, they are exchangeable).
However, in many real-world settings data are not exchangeable: for
example, time series data usually cannot be assumed to be exchangeable
due to temporal dependence. A recent line of research examines the
problem of generating prediction intervals for observations that are
observed online (that is, one at a time) and for which exchangeability
is not assumed to hold (Gibbs and Candes 2021; Zaffran et al. 2022;
Gibbs and Candès 2022; Bhatnagar et al. 2023). The methods from this
literature, which we refer to generally as \emph{Adaptive Conformal
Inference} (ACI) algorithms, work by adaptively adjusting the width of
the generated prediction intervals in response to the observed data.

Informally, suppose a sequence of outcomes \(y_t \in \mathbb{R}\),
\(t = 1, \dots, T\) are observed one at a time. Before seeing each
observation, we have at our disposal a point prediction
\(\hat{\mu}_t \in \mathbb{R}\) that can be generated by any method. Our
goal is to find an algorithm for producing prediction intervals
\([\ell_t, u_t]\), \(\ell_t \leq u_t\) such that, in the long run, the
observations \(y_t\) fall within the corresponding prediction intervals
roughly \(\alpha \times 100\%\) of the time: that is,
\(\lim_{T \to \infty} \sfrac{1}{T} \sum_{t=1}^T \mathbb{I}\{ y_t \in [\ell_t, u_t] \} = \alpha\).
The original ACI algorithm (Gibbs and Candes 2021) is based on a simple
idea: if the previous prediction interval at time \((t-1)\) did not
cover the true observation, then the next prediction interval at time
\(t\) is made slightly wider. Conversely, if the previous prediction
interval did include the observation, then the next prediction interval
is made slightly narrower. It can be shown that this procedure yields
prediction intervals that in the long run cover the true observations
the desired proportion of the time.

The main tuning parameter of the original ACI algorithm is a learning
rate that controls how fast prediction interval width changes. If the
learning rate is too low, then the prediction intervals will not be able
to adapt fast enough to shifts in the data generating distribution; if
it is too large, then the intervals will oscillate widely. The critical
dependence of the original ACI algorithm on proper choice of its
learning rate spurred subsequent research into meta-algorithms that
learn the correct learning rate (or an analogue thereof) in various
ways, typically drawing on approaches from the online learning
literature. In this paper, we present four such algorithms: Aggregated
ACI (AgACI, Zaffran et al. 2022), Fully Adaptive Conformal Inference
(FACI, Gibbs and Candès 2022), Scale-Free Online Gradient Descent
(SF-OGD, Bhatnagar et al. 2023), and Strongly Adaptive Online Conformal
Prediction (SAOCP, Bhatnagar et al. 2023). We note that the adaption of
conformal inference techniques is an active area of research and the
algorithms we focus on in this work are not exhaustive; see among others
Feldman et al. (2023), Bastani et al. (2022), Xu and Xie (2021), and Xu
and Xie (2023).

Our primary practical contribution is an implementation of each
algorithm in an open source \texttt{R} package,
\texttt{AdaptiveConformal}, which is available at
\url{https://github.com/herbps10/AdaptiveConformal}. The package also
includes routines for visualization and summary of the prediction
intervals. We note that Python versions of several algorithms were also
made available by Zaffran et al. (2022) and Bhatnagar et al. (2023), but
to our knowledge this is the first package implementing them in
\texttt{R}. In addition, several \texttt{R} packages exist for conformal
inference in other contexts, including \texttt{conformalInference}
focusing on regression (Tibshirani et al. 2019),
\texttt{conformalInference.fd}, with methods for functional responses
(Diquigiovanni et al. 2022), and \texttt{cfcausal} for causal inference
related functionals (Lei and Candès 2021). Our second practical
contribution is to compare the performance of the algorithms in
simulation studies and in a case study generating prediction intervals
for influenza incidence in the United States based on black-box point
forecasts.

The rest of the paper unfolds as follows. In Section~\ref{sec-theory},
we present a unified theoretical framework for analyzing the ACI
algorithms based on the online learning paradigm. In
Section~\ref{sec-algorithms} we provide descriptions of each algorithm
along with their known theoretical properties. In
Section~\ref{sec-simulations} we compare the performance of the
algorithms in several simulation studies. Section~\ref{sec-case-study}
gives a case study based on forecasting influenza in the United States.
Finally, Section~\ref{sec-discussion} provides a discussion and ideas
for future research in this rapidly expanding field.

\hypertarget{sec-theory}{%
\section{Theoretical Framework}\label{sec-theory}}

\emph{Notation}: for any integer \(N \geq 1\) let
\(\llbracket N \rrbracket := \{ 1, \dots, N \}\). Let \(\mathbb{I}\) be
the indicator function. Let \(\nabla f\) denote the gradient
(subgradient) of the differentiable (convex) function \(f\).

We consider an online learning scenario in which we gain access to a
sequence of observations \((y_t)_{t \geq 1}\) one at a time (see
Cesa-Bianchi and Lugosi (2006) for an comprehensive account of online
learning theory). Fix \(\alpha \in (0, 1)\) to be the target empirical
coverage of the prediction intervals. The goal is to output at time
\(t\) a prediction interval for the unseen observation \(y_{t}\), with
the prediction interval generated by an \emph{interval construction
function} \(\widehat{C}_{t}\). Formally, let \(\widehat{C}_t\) be a
function that takes as input a parameter \(\theta_t \in \mathbb{R}\) and
outputs a closed prediction interval \([\ell_t, u_t]\). The interval
construction function must be nested: if \(\theta^\prime > \theta\),
then \(\widehat{C}_t(\theta) \subseteq \widehat{C}_t(\theta^\prime)\).
In words, larger values of \(\theta\) imply wider prediction intervals.
The interval constructor is indexed by \(t\) to emphasize that it may
use other information at each time point, such as a point prediction
\(\hat{\mu}_t \in \mathbb{R}\). We make no restrictions on how this
external information is generated.

Define
\(r_t := \inf\{\theta \in \mathbb{R} : \mathbb{I}(y_t \in \widehat{C}_t(\theta)) \}\)
to be the \emph{radius} at time \(t\). The radius is the smallest
possible \(\theta\) such that the prediction interval covers the
observation \(y_t\). A key assumption for the theoretical analysis of
several of the algorithms is that the radii are bounded:

\textbf{Assumption}: there exists a \(D > 0\) such that \(r_t < D\) for
all \(t\).

Next, we describe two existing definitions of interval construction
functions.

\hypertarget{linear-intervals}{%
\subsection{Linear Intervals}\label{linear-intervals}}

A simple method for forming the prediction intervals is to use the
parameter \(\theta_t\) to directly define the width of the interval.
Suppose that at each time \(t\) we have access to a point prediction
\(\hat{\mu}_t \in \mathbb{R}\). Then we can form a symmetric prediction
interval around the point estimate using \[
\begin{aligned}
  \theta \mapsto \widehat{C}_t(\theta) := [\hat{\mu}_t - \theta, \hat{\mu}_t + \theta].
\end{aligned}
\] We refer to this as the \emph{linear interval constructor}. Note that
in this case, the radius is simply the absolute residual
\(r_t = |\hat{\mu}_t - y_t|\).

\hypertarget{quantile-intervals}{%
\subsection{Quantile Intervals}\label{quantile-intervals}}

The original ACI paper proposed constructing intervals based on the
previously observed residuals (Gibbs and Candes 2021). Let
\(S : \mathbb{R}^2 \to \mathbb{R}\) be a function called a
\emph{nonconformity score}. A popular choice of nonconformity score is
the absolute residual: \((\mu, y) \mapsto S(\mu, y):= |\mu - y|\). Let
\(s_t := S(\hat{\mu}_t, y_t)\) be the nonconformity score of the
\(t\)th-observation. The quantile interval construction function is then
given by \[
\begin{aligned}
  \widehat{C}_t(\theta_t) := [\hat{\mu}_t - \mathrm{Quantile}(\theta, \{ s_1, \dots, s_{t-1} \}), \hat{\mu}_t + \mathrm{Quantile}(\theta, \{ s_1, \dots, s_{t-1} \})]
\end{aligned}
\] where \(\mathrm{Quantile}(\theta, A)\) denotes the empirical
\(\theta\)-quantile of the elements in the set \(A\). Note that
\(\widehat{C}_t\) is indeed nested in \(\theta_t\) because the Quantile
function is non-decreasing in \(\theta\).

Our proposed \texttt{AdaptiveConformal} package takes the absolute
residual as the default nonconformity score, although the user may also
specify any custom nonconformity score by supplying it as an \texttt{R}
function.

\hypertarget{online-learning-framework}{%
\subsection{Online Learning Framework}\label{online-learning-framework}}

We now introduce a loss function that defines the quality of a
prediction interval with respect to a realized observation. Define the
\emph{pinball loss} \(L^\alpha\) as \[
(\theta, r) \mapsto L^\alpha(\theta, r) := \begin{cases}
  \alpha(\theta - r), & \theta \geq r \\
  (1-\alpha)(r - \theta), & \theta < r.
\end{cases}
\] The way in which the algorithm gains access to the data and incurs
losses is as follows:

\begin{itemize}
\tightlist
\item
  Sequentially, for \(t = 1, \dots, T\):

  \begin{itemize}
  \tightlist
  \item
    Predict radius \(\theta_t\) and form prediction interval
    \(\widehat{C}_t(\theta_t)\).
  \item
    Observe true outcome \(y_t\) and calculate radius \(r_t\).
  \item
    Record
    \(\mathrm{err}_t := \mathbb{I}[y_t \not\in \widehat{C}_t(\theta_t)]\).
  \item
    Incur loss \(L^\alpha(\theta_t, r_t)\).
  \end{itemize}
\end{itemize}

This iterative procedure is at the core of the online learning
theoretical framework in which theoretical results have been derived.

\hypertarget{assessing-aci-algorithms}{%
\subsection{Assessing ACI algorithms}\label{assessing-aci-algorithms}}

There are two different perspectives we can take in measuring the
quality of an ACI algorithm that generates a sequence
\((\theta_t)_{t \in \llbracket T \rrbracket}\). First, we could look at
how close the empirical coverage of the generated prediction intervals
is to the desired coverage level \(\alpha\). Formally, define the
empirical coverage as the proportion of observations that fell within
the corresponding prediction interval:
\(\mathrm{EmpCov}(T) := \frac{1}{T} \sum_{t=1}^T (1 - \mathrm{err}_t)\).
The coverage error is then given by \[
\begin{aligned}
  \mathrm{CovErr}(T) := \mathrm{EmpCov}(T) - \alpha.
\end{aligned}
\] The second perspective is to look at how well the algorithm controls
the incurred pinball losses. Following the classical framework from the
online learning literature, we define the \emph{regret} as the
difference between the cumulative loss yielded by a sequence
\((\theta_t)_{t \in \llbracket T \rrbracket}\) versus the cumulative
loss of the best possible fixed choice: \[
\begin{aligned}
  \mathrm{Reg}(T) := \sum_{t=1}^T L^\alpha(\theta_t, r_t) - \inf_{\theta^* \in \mathbb{R}} \sum_{t=1}^T L^\alpha(\theta^*, r_t).
\end{aligned}
\] In settings of distribution shift, it may not be appropriate to
compare the cumulative loss of an algorithm to a fixed competitor. As
such, stronger notions of regret have been defined. The \emph{strongly
adaptive regret} is the largest regret over any subperiod of length
\(k \in \llbracket T \rrbracket\): \[
\begin{aligned}
  \mathrm{SAReg}(T, m) := \max_{[\tau, \tau + m - 1] \subseteq \llbracket T \rrbracket} \left( \sum_{t=\tau}^{\tau + m - 1} L^{\alpha}(\theta_t, r_t) - \inf_{\theta^* \in \mathbb{R}} \sum_{t=\tau}^{\tau + m - 1} L^\alpha(\theta^*, r_t) \right).
\end{aligned}
\] Both ways of evaluating ACI methods are important because targeting
only one or the other can lead to algorithms that yield prediction
intervals that are not practically useful. As a simple pathological
example of only targeting the coverage error, suppose we wish to
generate \(\alpha = 50\%\) prediction intervals. We could choose to
alternate \(\theta\) between 0 and \(\infty\), such that
\(\mathrm{err}_t\) alternates between 0 and 1. The empirical coverage
would then trivially converge to the desired level of 50\%. However, the
same algorithm would yield infinite regret (see Bhatnagar et al. (2023)
for a more in-depth example of an scenario in which coverage is optimal
but the regret grows linearly). On the other hand, an algorithm that has
arbitrarily small regret may not yield good empirical coverage. Suppose
the observations and point predictions are constant: \(y_t = 1\) and
\(\hat{\mu}_t = 0\) for all \(t \geq 1\). Consider a simple class of
algorithms that outputs constantly \(\theta_t = \theta'\) for some
\(\theta' < 1\). With the linear interval construction function, the
prediction intervals are then
\(\widehat{C}_t(\theta_t) = [-\theta', \theta']\). The regret is given
by \(\mathrm{Reg}(T) = 2T\alpha(1-\theta')\), which approaches zero as
\(\theta'\) approaches 1. The empirical coverage is, however, always
zero. In other words, the regret can be arbitrarily close to zero while
at the same time the empirical coverage does not approach the desired
level.

These simple examples illustrate that, unfortunately, bounds on the
coverage error and bounds on the regret are not in general
interchangeable. It is possible, however, to show equivalencies by
either (1) making distributional assumptions on the data or (2) using
additional information about how the algorithm produces the sequence
\((\theta_t)_{t \in \llbracket T \rrbracket}\) (Bhatnagar et al. 2023).

It may also be informative to summarize a set of prediction intervals in
ways beyond their coverage error or their regret. A common metric for
prediction intervals is the \emph{mean interval width}: \[
\begin{aligned}
  \mathrm{MeanWidth}(T) := \frac{1}{T} \sum_{t=1}^T w_t,
\end{aligned}
\] where \(w_t := u_t - \ell_t\) is the interval width at time \(t\).

Finally, we introduce a metric that is intended to capture pathological
behavior that can arise with ACI algorithms where the prediction
intervals oscillate between being extremely narrow and extremely wide.
Define the \emph{path length} of prediction intervals generated by an
ACI algorithm as \[
\begin{aligned}
  \mathrm{PathLength}(T) := \sum_{t=2}^T |w_t - w_{t-1}|.
\end{aligned}
\] A high path length indicates that the prediction intervals were
variable over time, and a low path length indicates the prediction
intervals were stable.

\hypertarget{sec-algorithms}{%
\section{Algorithms}\label{sec-algorithms}}

\hypertarget{tbl-aci}{}
\begin{longtable}[]{@{}
  >{\raggedright\arraybackslash}p{(\columnwidth - 4\tabcolsep) * \real{0.3971}}
  >{\raggedright\arraybackslash}p{(\columnwidth - 4\tabcolsep) * \real{0.4485}}
  >{\raggedright\arraybackslash}p{(\columnwidth - 4\tabcolsep) * \real{0.1544}}@{}}
\caption{\label{tbl-aci}Summary of ACI algorithms}\tabularnewline
\toprule\noalign{}
\begin{minipage}[b]{\linewidth}\raggedright
Algorithm
\end{minipage} & \begin{minipage}[b]{\linewidth}\raggedright
Tuning Parameters
\end{minipage} & \begin{minipage}[b]{\linewidth}\raggedright
Original interval constructor
\end{minipage} \\
\midrule\noalign{}
\endfirsthead
\toprule\noalign{}
\begin{minipage}[b]{\linewidth}\raggedright
Algorithm
\end{minipage} & \begin{minipage}[b]{\linewidth}\raggedright
Tuning Parameters
\end{minipage} & \begin{minipage}[b]{\linewidth}\raggedright
Original interval constructor
\end{minipage} \\
\midrule\noalign{}
\endhead
\bottomrule\noalign{}
\endlastfoot
Adaptive Conformal Inference (ACI) & Learning rate \(\gamma\) &
Quantile \\
Aggregated Adaptive Conformal Inference (AgACI) & Candidate learning
rates \((\gamma_k)_{1 \leq k \leq K}\) & Quantile \\
Fully Adaptive Conformal Inference (FACI) & Candidate learning rates
\((\gamma_k)_{1 \leq k \leq K}\) & Quantile \\
Scale-Free Online Gradient Descent (SF-OGD) & Learning rate \(\gamma\)
or maximum radius \(D\) & Linear \\
Strongly Adaptive Online Conformal Prediction (SAOCP) & Learning rate
\(\gamma\), lifetime multiplier \(g\) & Linear \\
\end{longtable}

As a simple running example to illustrate each algorithm, we simulate
independently \(T = 500\) values \(y_1, \dots, y_T\) following \[
y_t \sim N(0, 0.2^2), \quad t \in \llbracket T \rrbracket.
\] For demonstration purposes we assume we have access to unbiased
predictions \(\hat{\mu}_t = 0\) for all
\(t \in \llbracket T \rrbracket\). Throughout we set the target
empirical coverage to \(\alpha = 0.8\).

\hypertarget{adaptive-conformal-inference-aci}{%
\subsection{Adaptive Conformal Inference
(ACI)}\label{adaptive-conformal-inference-aci}}

\begin{algorithm}[htb!]
\caption{Adaptive Conformal Inference}
\label{algo-aci}
\begin{algorithmic}[1]
\State \textbf{Input:} starting value $\theta_1$, learning rate $\gamma > 0$.
\For{$t = 1, 2, \dots, T$}
  \State \textbf{Output:} prediction interval $\widehat{C}_t(\theta_t)$.
  \State Observe $y_t$.
  \State Evaluate $\mathrm{err}_t = \mathbb{I}[y_t \not\in \widehat{C}_t(\theta_t)]$.
  \State Update $\theta_{t+1} = \theta_t + \gamma (\mathrm{err}_t - (1 - \alpha))$.
\EndFor
\end{algorithmic}
\end{algorithm}

The original ACI algorithm (Gibbs and Candes (2021);
 Algorithm~\ref{algo-aci} ) adaptively adjusts the width of the
prediction intervals in response to the observations. The updating rule
for the estimated radius can be derived as an online subgradient descent
scheme. The subgradient of the pinball loss function with respect to
\(\theta\) is given by \[
\begin{aligned}
  \nabla L^\alpha(\theta, r) &= \begin{cases}
    \{ -\alpha \}, &\theta < r, \\
    \{ 1 - \alpha \}, & \theta > r, \\
    [-\alpha, 1 - \alpha], &\theta = r
  \end{cases} \\
\end{aligned}
\] It follows that, for all \(\theta_t \in \mathbb{R}\) and
\(r_t \in \mathbb{R}\), \[
1 - \alpha - \mathrm{err}_t \in \nabla L^\alpha(\theta_t, r_t).
\] This leads to the following update rule for \(\theta\) based on
subgradient descent: \[
\begin{aligned}
  \theta_{t+1} = \theta_{t} + \gamma (\mathrm{err}_t - (1 - \alpha)),
\end{aligned}
\] where \(\gamma > 0\) is a user-specified learning rate. For
intuition, note that if \(y_t\) fell outside of the prediction interval
at time \(t\) (\(\mathrm{err}_t = 1\)) then the next interval is widened
(\(\theta_{t+1} = \theta_t + \gamma \alpha\)). On the contrary, if
\(y_t\) fell within the interval (\(\mathrm{err}_t = 0\)) then the next
interval is shortened
(\(\theta_{t+1} = \theta_t - \gamma(1 - \alpha)\)). The learning rate
\(\gamma\) controls how fast the width of the prediction intervals
changes in response to the data.

\hypertarget{theoretical-guarantees}{%
\subsubsection{Theoretical Guarantees}\label{theoretical-guarantees}}

The ACI algorithm has the following finite sample bound on the coverage
error (Gibbs and Candes 2021). For all \(\gamma > 0\), \[
|\mathrm{CovErr}(T)| \leq \frac{D + \gamma}{\gamma T}.
\] This result was originally shown for ACI with the choice of the
quantile interval constructor, although it can also be extended to other
interval constructors Feldman et al. (2023). In addition, standard
results for online subgradient descent yield the following regret bound
with the use of the linear interval constructor, assuming that the true
radii are bounded by \(D\): \[
\mathrm{Reg}(T) \leq \mathcal{O}(D^2 / \gamma + \gamma T) \leq \mathcal{O}(D \sqrt{T}),
\] where the second inequality follows if the optimal choice of
\(\gamma = D/\sqrt{T}\) is used (Bhatnagar et al. 2023). Taken together,
these theoretical results imply that while the coverage error is
guaranteed to converge to zero for any choice of \(\gamma\), achieving
sublinear regret requires choosing \(\gamma\) more carefully. This
highlights the importance of both ways of assessing ACI algorithms: if
we only focused on controlling the coverage error, we might not achieve
optimal control of regret, leading to intervals that are not practically
useful.

\hypertarget{tuning-parameters}{%
\subsubsection{Tuning Parameters}\label{tuning-parameters}}

\begin{figure}

{\centering \includegraphics{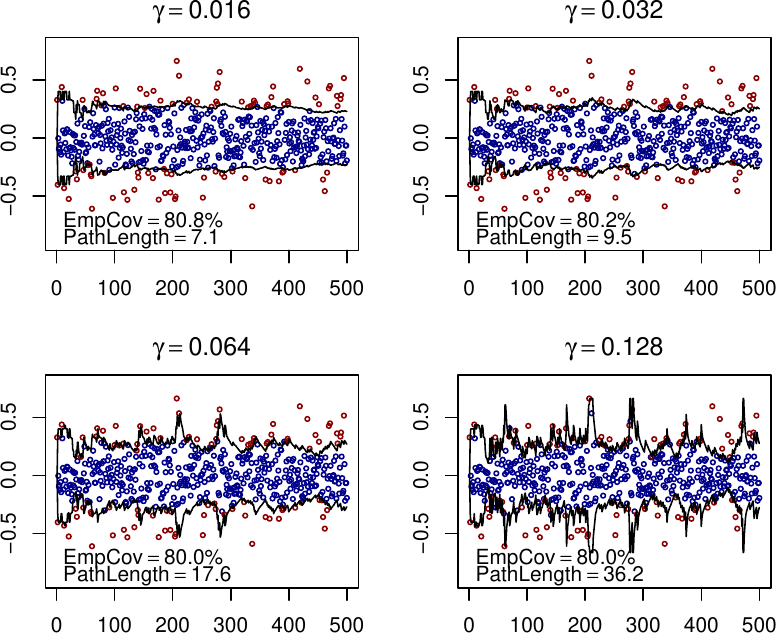}

}

\caption{\label{fig-aci}Example 80\% prediction intervals from the ACI
algorithm for different choices of learning rate \(\gamma\) and with
\(\theta_1 = 0.8\). Blue and red points are observations that fell
inside and outside the prediction intervals, respectively.}

\end{figure}

Therefore, the main tuning parameter is the learning rate \(\gamma\).
The theoretical bounds on the coverage error suggest setting a large
\(\gamma\) such that the coverage error decays quickly in \(T\);
however, in practice and setting \(\gamma\) too large will lead to
intervals with large oscillations as seen in Figure~\ref{fig-aci}. This
is quantified in the path length, which increases significantly as
\(\gamma\) increases, even though the empirical coverage remains near
the desired value of 80\%. On the other hand, setting \(\gamma\) too
small will lead to intervals that do not adapt fast enough to
distribution shifts. Thus, choosing a good value for \(\gamma\) is
essential. However, the optimal choice \(\gamma = D / \sqrt{T}\) cannot
be used directly in practice unless the time series length \(T\) is
fixed in advance, or the so called ``doubling trick'' is used to relax
the need to know \(T\) in advance (Cesa-Bianchi and Lugosi 2006).

The theoretical results guaranteeing the performance of the ACI
algorithm do not depend on the choice of starting value \(\theta_1\),
and thus in practice any value can be chosen. Indeed, the effect of the
choice of \(\theta_1\) decays over time as a function of the chosen
learning rate. In practice, substantive prior information can be used to
pick a reasonable starting value. By default, the
\texttt{AdaptiveConformal} package sets \(\theta_1 = \alpha\) when the
quantile interval predictor is used, and \(\theta_1 = 0\) otherwise,
although in both cases the user can supply their own starting value. The
behavior of the early prediction intervals in the examples
(Figure~\ref{fig-aci}) is driven by the small number of residuals
available, which makes the output of the quantile interval constructor
sensitive to small changes in \(\theta\). In practice, a warm-up period
can be used before starting to produce prediction intervals so that the
quantiles of the residuals are more stable.

\hypertarget{aggregated-adaptive-conformal-inference-agaci}{%
\subsection{Aggregated Adaptive Conformal Inference
(AgACI)}\label{aggregated-adaptive-conformal-inference-agaci}}

\begin{algorithm}[htb!]
\caption{Aggregated Adaptive Conformal Inference}
\label{algo-agaci}
\begin{algorithmic}[1]
  \State \textbf{Input:} candidate learning rates $(\gamma_k)_{1 \leq k \leq K }$, starting value $\theta_1$.
  \State Initialize lower and upper BOA algorithms $\mathcal{B}^\ell := \texttt{BOA}(\alpha \leftarrow (1 - \alpha) / 2)$ and $\mathcal{B}^u := \texttt{BOA}(\alpha \leftarrow (1 - (1 - \alpha)/2))$.
  \For{$k = 1, \dots, K$}
    \State Initialize ACI $\mathcal{A}_k = \texttt{ACI}(\alpha \leftarrow \alpha, \gamma \leftarrow \gamma_k, \theta_1 \leftarrow \theta_1)$.
  \EndFor
  \For{$t = 1, 2, \dots, T$}
    \For{$k = 1, \dots, K$}
      \State Retrieve candidate prediction interval $[\ell^k_{t}, u^k_{t}]$ from $\mathcal{A}_k$.
    \EndFor
    \State Compute aggregated lower bound $\tilde{\ell}_t := \mathcal{B}^\ell((\ell^k_t : k \in \{ 1, \dots, K \}))$.
    \State Compute aggregated upper bound $\tilde{u}_t := \mathcal{B}^u((u^k_t : k \in \{ 1, \dots, K \}))$.
    \State \textbf{Output:} prediction interval $[\tilde{\ell}_t, \tilde{u}_t]$.
    \State Observe $y_t$.
    \For{$k = 1, \dots, K$}
      \State Update $\mathcal{A}_k$ with observation $y_t$.
    \EndFor
    \State Update $\mathcal{B}^\ell$ with observed outcome $y_t$. 
    \State Update $\mathcal{B}^u$ with observed outcome $y_t$.
  \EndFor
\end{algorithmic}
\end{algorithm}

The Aggregated ACI (AgACI;  Algorithm~\ref{algo-agaci} ) algorithm
solves the problem of choosing a learning rate for ACI by running
multiple copies of the algorithm with different learning rates, and then
separately combining the lower and upper interval bounds using an online
aggregation of experts algorithm (Zaffran et al. 2022). That is, one
aggregation algorithm seeks to estimate the lower \((1-\alpha)/2\)
quantile, and the other seeks to estimate the upper
\(1 - (1 - \alpha) / 2\) quantile. Zaffran et al. (2022) experimented
with multiple online aggregation algorithms, and found that they yielded
similar results. Thus, we follow their lead in using the Bernstein
Online Aggregation (BOA) algorithm as implemented in the \texttt{opera}
\texttt{R} package (Wintenberger 2017; Gaillard et al. 2023). BOA is an
online algorithm that forms predictions for the lower (or upper)
prediction interval bound as a weighted mean of the candidate ACI
prediction interval lower (upper) bound, where the weights are
determined by each candidate's past performance with respect to the
quantile loss. As a consequence, the prediction intervals generated by
AgACI are not necessarily symmetric around the point prediction, as the
weights for the lower and upper bounds are separate.

\hypertarget{theoretical-gaurantees}{%
\subsubsection{Theoretical Gaurantees}\label{theoretical-gaurantees}}

AgACI departs from our main theoretical framework in that it does not
yield a sequence \((\theta_t)_{t \in \llbracket T \rrbracket}\) whose
elements yield prediction intervals via a set construction function
\(\widehat{C}_t\). Rather, the upper and lower interval bounds from a
set of candidate ACI algorithms are aggregated separately. Thus,
theoretical results such as regret bounds similar to those for the other
algorithms are not available. It would be possible, however, to
establish regret bounds for the pinball loss applied separately to the
lower and upper bounds of the prediction intervals. It is unclear,
however, how to convert such regret bounds into a coverage bound.

\hypertarget{tuning-parameters-1}{%
\subsubsection{Tuning Parameters}\label{tuning-parameters-1}}

\begin{figure}

{\centering \includegraphics{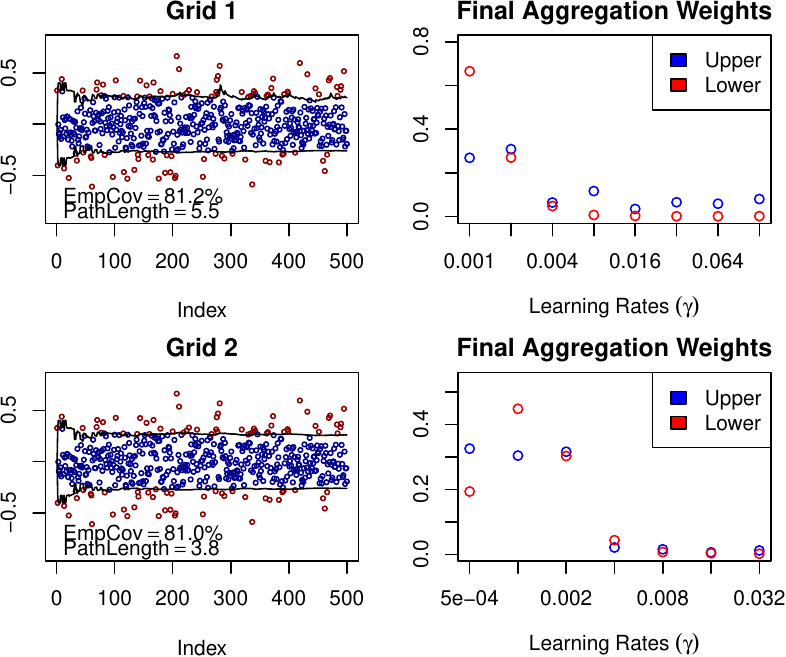}

}

\caption{\label{fig-agaci}Example 80\% prediction intervals from the
AgACI algorithm with starting values \(\theta_1 = 0.8\) and two
different learning rate grids, where the second grid is double the size
of the first. In the left column, blue and red points are observations
that fell inside and outside the prediction intervals, respectively.}

\end{figure}

The main tuning parameter for AgACI is the set of candidate learning
rates. Beyond necessitating additional computational time, there is no
drawback to having a large grid. As a default,
\texttt{AdaptiveConformal} uses learning rates
\(\gamma \in \{ 0.001, 0.002, 0.004, 0.008, 0.016, 0.032, 0.064, 0.128 \}\).
As a basic check, we can look at the weights assigned to each of the
learning rates. If large weights are given to the smallest (largest)
learning rates, it is a sign that smaller (or larger) learning rates may
perform well. In addition each of the candidate ACI algorithms requires
a starting value, which can be set to any value as discussed in the ACI
section. Figure~\ref{fig-agaci} illustrates AgACI applied to the running
example with two sets of learning grids. The first grid is the default,
and the second grid includes the additional value
\(\gamma = \{ 0.0005 \}\). For the first grid, we can see that for the
lower bound AgACI assigns high weight to the lowest learning rate
(\(\gamma = 0.001\)). Based on this observation, as a sensitivity check,
we reran the algorithm with the second learning grid, which yields
weights that are less concentrated on a single learning rate. The output
prediction intervals are similar, although slightly smoother in the
second grid, reflecting the lack of distributional shift in the
simulated data.

\hypertarget{fully-adaptive-conformal-inference-faci}{%
\subsection{Fully Adaptive Conformal Inference
(FACI)}\label{fully-adaptive-conformal-inference-faci}}

\begin{algorithm}[htb!]
\caption{Fully Adaptive Conformal Inference}
\label{algo-faci}
\begin{algorithmic}[1]
\State \textbf{Input:} starting value $\theta_1$, candidate learning rates $(\gamma_k)_{1 \leq k \leq K }$, parameters $\sigma, \eta$.
\For{$k = 1, \dots, K$}
  \State Initialize expert $\mathcal{A}_k = \texttt{ACI}(\alpha \leftarrow \alpha, \gamma \leftarrow \gamma_k, \theta_1 \leftarrow \theta_1)$.
\EndFor
\For{$t = 1, 2, \dots, T$}
  \State Define $p_t^k := w_t^k / \sum_{i=1}^K w_t^i$, for all $1 \leq k \leq K$.
  \State Set $\theta_t = \sum_{k=1}^K \theta_t^k p_t^k$.
  \State \textbf{Output:} prediction interval $\widehat{C}_t(\theta_t)$.
  \State Observe $y_t$ and compute $r_t$.
  \State $\bar{w}_{t}^k \gets w_t^k \exp(-\eta L^\alpha(\theta_t^k, r_t))$, for all $1 \leq k \leq K$.
  \State $\bar{W}_t \gets \sum_{i=1}^K \bar{w}_t^i$.
  \State $w_{t+1}^k \gets (1 - \sigma) \bar{w}_t^k + \bar{W}_t \sigma / K$.
  \State Set $\mathrm{err}_t := \mathbb{I}[y_t \not\in \widehat{C}_t(\theta_t)]$.
  \For{$k = 1, \dots, K$}
    \State Update ACI $\mathcal{A}_k$ with $y_t$ and obtain $\theta_{t+1}^k$.
  \EndFor
\EndFor
\end{algorithmic}
\end{algorithm}

The Fully Adaptive Conformal Inference (FACI;
 Algorithm~\ref{algo-faci} ) algorithm was developed by the authors of
the original ACI algorithm in part to address the issue of how to choose
the learning rate parameter \(\gamma\). In this respect the goal of the
algorithm is similar to that of AgACI, although it is achieved slightly
differently. FACI also aggregates predictions from multiple copies of
ACI run with different learning rates, but differs in that it directly
aggregates the estimated radii emitted from each algorithm based on
their pinball loss (Gibbs and Candès 2022) using an exponential
reweighting scheme (Gradu, Hazan, and Minasyan 2023). As opposed to
AgACI, this construction allows for more straightforward development of
theoretical guarantees on the algorithm's performance, because the upper
and lower bounds of the intervals are not aggregated separately.

\hypertarget{theoretical-guarantees-1}{%
\subsubsection{Theoretical Guarantees}\label{theoretical-guarantees-1}}

FACI was originally proposed with the choice of the quantile interval
constructor. FACI has the following strongly-adaptive regret bound
(Bhatnagar et al. 2023): for all \(\gamma > 0\) and subperiod lengths
\(m\), \[
\begin{aligned}
  \mathrm{SAReg}(T, m) \leq \widetilde{\mathcal{O}}(D^2 / \gamma + \gamma m).
\end{aligned}
\] If \(m\) is fixed a-priori, then choosing \(\gamma = D/\sqrt{m}\)
yields a strongly adaptive regret bound of order
\(\widetilde{\mathcal{O}}(D \sqrt{m})\) (for a single choice of \(m\)).
Practically, this result implies that, if we know in advance the time
length for which we would like to control the regret, it is possible to
choose an optimal tuning parameter value. However, we cannot control the
regret simultaneously for all possible time lengths.

To establish a bound on the coverage error, the authors investigated a
slightly modified version of FACI in which \(\theta_t\) is chosen
randomly from the candidate \(\theta_{t_k}\) with weights given by
\(p_{t,k}\), instead of taking a weighted average. This is a common
trick used in the literature as it facilitates theoretical analysis. In
practice, the authors comment that this randomized version of FACI and
the deterministic version lead to very similar results. The coverage
error result also assumes that the hyperparameters can change over time:
that is, we have \(t\)-specific \(\eta_{t}\) and \(\sigma_t\), rather
than fixed \(\eta\) and \(\sigma\). The coverage error then has the
following bound (Gibbs and Candès 2022), where \(\gamma_{\min}\) and
\(\gamma_{\max}\) are the smallest and largest learning rates in the
grid, respectively: \[
|\mathrm{CovErr}(T)| \leq \frac{1 + 2\gamma_{\max}}{T \gamma_{\min}} + \frac{(1 + 2\gamma_{\max})^2}{\gamma_{\min}} \exp(\eta_t(1 + 2\gamma_{\max})) \frac{1}{T}\sum_{t=1}^T \eta_t + 2 \frac{1+\gamma_{\max}}{\gamma_{\min}} \frac{1}{T} \sum_{t=1}^T \sigma_t.
\] Thus, if \(\eta_t\) and \(\sigma_t\) both converge to zero as
\(t \to \infty\), then the coverage error will also converge to zero. In
addition, under mild distributional assumptions the authors provide a
type of short-term coverage error bound for arbitrary time spans, for
which we refer to (Gibbs and Candès 2022).

We note one additional result established by Gibbs and Candès (2022) on
a slightly different regret bound in terms of the pinball loss, as it
informs the choice of tuning parameters. Let
\(\gamma_{\mathrm{max}} = \max_{1 \leq k \leq K} \gamma_k\) be the
largest learning rate in the grid and assume that
\(\gamma_1 < \gamma_2 < \cdots < \gamma_K\) with
\(\gamma_{k+1}/\gamma \leq 2\) for all \(1 \leq k < K\). Then, for any
interval \(I = [r, s] \subseteq \llbracket T \rrbracket\) and any
sequence \(\theta_r^*, \dots, \theta_s^*\), under the assumption that
\(\gamma_k \geq \sqrt{1 + 1 / |I|}\), \[
\begin{aligned}
  \frac{1}{|I|} \sum_{t=r}^s \mathbb{E}[L^\alpha(\theta_t, r_t)] - \frac{1}{|I|} \sum_{t=r}^s L^\alpha(\theta_t, \theta_t^*) \leq& \frac{\log(k / \sigma) + 2\sigma|I|}{\eta |I|} + \frac{\eta}{|I|} \sum_{t=r}^s \mathbb{E}[L^\alpha(\theta_t, r_t)^2] \\
  &+ 2\sqrt{3}(1 + \gamma_{\mathrm{max}})^2 \max\left\{ \sqrt{\frac{\sum_{t=r+1}^s |\theta_t^* - \theta_{t-1}^*| + 1}{|I|}}, \gamma_1 \right\},
\end{aligned}
\] where the expectation is over the randomness in the randomized
version of the algorithm. Here the time interval \(I\) (with length
\(|I|\)) is comparable to the time period length \(m\) for the strongly
adaptive regret. The parameter \(|I|\), the time interval of interest
for which we would like to control, can be chosen arbitrarily.

\hypertarget{tuning-parameters-2}{%
\subsubsection{Tuning parameters}\label{tuning-parameters-2}}

The recommended settings for the tuning parameters depend on choosing a
time interval length \(|I|\) for which we would like to control the
pinball loss. The choice of \(|I|\) can be chosen arbitrarily. For the
tuning parameter \(\sigma\), the authors suggest the optimal choice
\(\sigma = 1 / (2 |I|)\). Choosing \(\eta\) is more difficult. The
authors suggest the following choice for \(\eta\), which they show is
optimal if there is in fact no distribution shift: \[
\begin{aligned}
  \eta = \sqrt{\frac{3}{|I|}} \sqrt{\frac{\log(K \cdot |I|) + 2}{(\alpha)^2 (1 - \alpha)^3 + (1-\alpha)^2 \alpha^3 }}
\end{aligned}.
\] Note that this choice is optimal only for the quantile interval
constructor, for which \(\theta_t\) is a quantile of previous
nonconformity scores. As an alternative, the authors point out that
\(\eta\) can be learned in an online fashion using the update rule \[
\begin{aligned}
  \eta_t := \sqrt{\frac{\log(|I| K) + 2}{\sum_{s=t-|I|}^{t-1} L^\alpha(\theta_s, r_s)}}.
\end{aligned}
\] Both ways of choosing \(\eta\) led to very similar results in the
original author's empirical studies. In our proposed
\texttt{AdaptiveConformal} package, the first approach is used when the
quantile interval construction function is chosen, and the latter
approach for the linear interval construction function.

\begin{figure}

{\centering \includegraphics{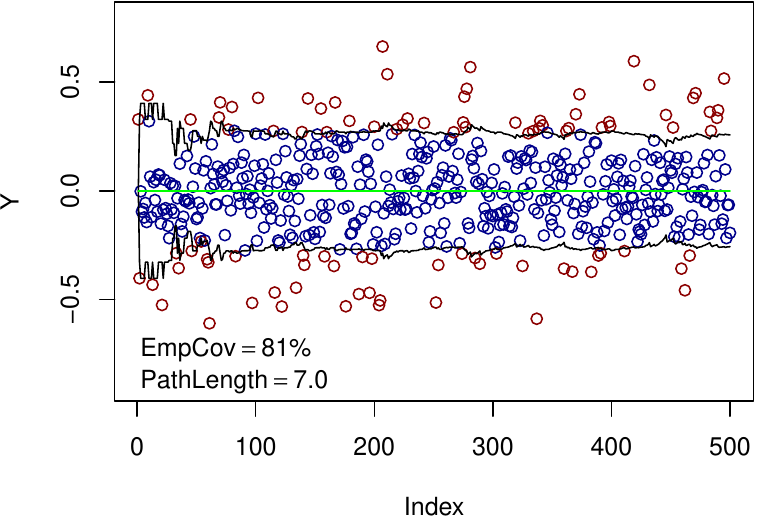}

}

\caption{\label{fig-faci-example}Example 80\% prediction intervals
generated by the FACI algorithm with starting values \(\theta_1 = 0.8\).
Blue and red points are observations that fell inside and outside the
prediction intervals, respectively.}

\end{figure}

Figure~\ref{fig-faci-example} illustrates FACI with the quantile
interval construction function and with the learning rate grid
\(\gamma \in \{ 0.001, 0.002, 0.004, 0.008, 0.016, 0.032, 0.064, 0.128 \}\).

\hypertarget{scale-free-online-gradient-descent-sf-ogd}{%
\subsection{Scale-Free Online Gradient Descent
(SF-OGD)}\label{scale-free-online-gradient-descent-sf-ogd}}

\begin{algorithm}[htb!]
\caption{Scale-Free Online Gradient Descent}
\label{algo-sfogd}
\begin{algorithmic}[1]
\State \textbf{Input:} starting value $\theta_1$, learning rate $\gamma > 0$.
\For{$t = 1, 2, \dots, T$}
  \State \textbf{Output:} prediction interval $\widehat{C}_t(\theta_t)$. 
  \State Observe $y_t$ and compute $r_t$. 
  \State Update $\theta_{t+1} = \theta_t - \gamma \frac{\nabla L^\alpha(\theta_t, r_t)}{\sqrt{\sum_{i=1}^t} \| \nabla L^\alpha(\theta_i, r_i) \|_2^2}$.
\EndFor
\end{algorithmic}
\end{algorithm}

Scale-Free Online Gradient Descent (SF-OGD;
 Algorithm~\ref{algo-sfogd} ) is a general algorithm for online learning
proposed by Orabona and Pál (2018). The algorithm updates \(\theta_t\)
with a gradient descent step where the learning rate adapts to the scale
of the previously observed gradients. SF-OGD was first used in the
context of ACI as a sub-algorithm for SAOCP (described in the next
section). However, it was found to have good performance by itself
(Bhatnagar et al. 2023) in real-world tasks, so we have made it
available in the package as a stand-alone algorithm.

\hypertarget{theoretical-guarantees-2}{%
\subsubsection{Theoretical Guarantees}\label{theoretical-guarantees-2}}

The SF-OGD algorithm with linear interval constructor has the following
regret bound, which is called an \emph{anytime regret bound} because it
holds for all \(t \in \llbracket T \rrbracket\) (Bhatnagar et al. 2023).
For any \(\gamma > 0\), \[
\begin{aligned}
  \mathrm{Reg}(t) \leq \mathcal{O}(D \sqrt{t}) \text{ for all } t \in \llbracket T \rrbracket.
\end{aligned}
\] A bound for the coverage error has also been established (Bhatnagar
et al. 2023). For any learning rate \(\gamma = \Theta(D)\) (where
\(\gamma = D / \sqrt{3}\) is optimal) and any starting value
\(\theta_1 \in [0, D]\), then it holds that for any \(T > 1\), \[
\begin{aligned}
  |\mathrm{CovErr}(T)| \leq \mathcal{O}\left( (1 - \alpha)^{-2} T^{-1/4} \log T \right).
\end{aligned}
\]

\hypertarget{tuning-parameters-3}{%
\subsubsection{Tuning parameters}\label{tuning-parameters-3}}

\begin{figure}

{\centering \includegraphics{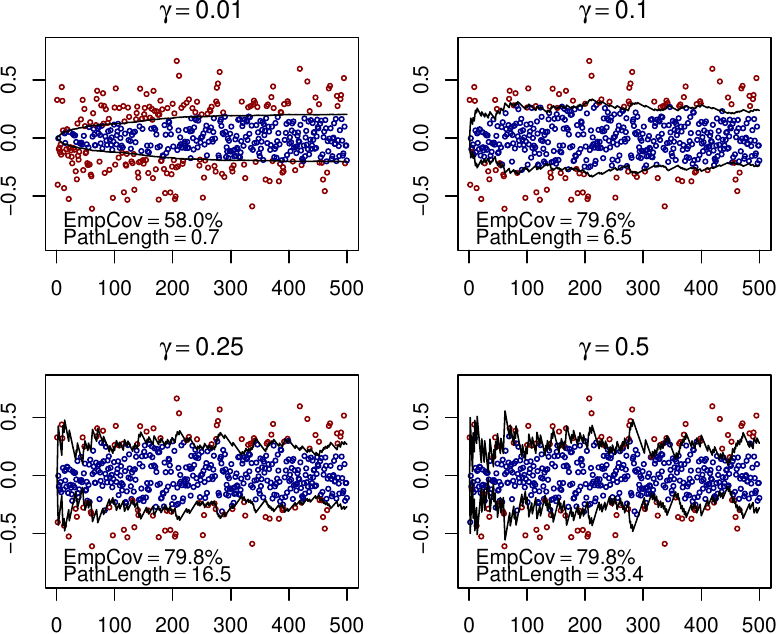}

}

\caption{\label{fig-sf-ogd}Example 80\% prediction intervals generated
by the SF-OGD algorithm with different values of the maximum radius
tuning parameter \(D\). Blue and red points are observations that fell
inside and outside the prediction intervals, respectively.}

\end{figure}

Figure~\ref{fig-sf-ogd} compares results for several choices of
\(\gamma\) to illustrate its effect. The optimal choice of learning rate
is \(\gamma = D / \sqrt{3}\), where \(D\) is the maximum possible
radius. When \(D\) is not known, it can be estimated by using an initial
subset of the time series as a calibration set and estimating \(D\) as
the maximum of the absolute residuals of the observations and the
predictions (Bhatnagar et al. 2023). Figure~\ref{fig-sf-ogd} illustrates
SF-OGD for several values of \(\gamma\).

\hypertarget{strongly-adaptive-online-conformal-prediction-saocp}{%
\subsection{Strongly Adaptive Online Conformal Prediction
(SAOCP)}\label{strongly-adaptive-online-conformal-prediction-saocp}}

\begin{algorithm}[htb!]
\caption{Strongly Adaptive Online Conformal Prediction}
\label{algo-saocp}
\begin{algorithmic}[1]
\State \textbf{Input:} initial value $\theta_0$, learning rate $\gamma > 0$.
\For{$t = 1, 2, \dots, T$}
  \State Initialize expert $\mathcal{A}_t = \texttt{SF-OGD}(\alpha \leftarrow \alpha, \gamma \leftarrow \gamma, \theta_1 \leftarrow \theta_{t-1})$, set weight $w_t^t = 0$.
  \State Compute active set $\mathrm{Active}(t) = \{ i \in \llbracket T \rrbracket : t - L(i) < i \leq t \}$ (see below for definition of $L(t)$).
  \State Compute prior probability $\pi_i \propto i^{-2} (1 + \lfloor \log_2 i \rfloor )^{-1} \mathbb{I}[i \in \mathrm{Active}(t)]$.
  \State Compute un-normalized probability $\hat{p}_i = \pi_i [w_{t,i}]_+$ for all $i \in \llbracket t \rrbracket$.
  \State Normalize $p = \hat{p} / \| \hat{p} \|_1 \in \Delta^t$ if $\| \hat{p} \|_1 > 0$, else $p = \pi$.
  \State Set $\theta_t = \sum_{i \in \mathrm{Active}(t)} p_i \theta_t^i$ (for $t \geq 2$), and $\theta_t = 0$ for $t = 1$.
  \State \textbf{Output:} prediction set $\widehat{C}_t(\theta_t)$.
  \State Observe $y_t$ and compute $r_t$. 
  \For{$i \in \mathrm{Active}(t)$}
    \State Update expert $\mathcal{A}_t$ with $y_t$ and obtain $\theta_{t+1}^i$.
    \State Compute $g_t^i = \begin{cases}
      \frac{1}{D}\left(L^\alpha(\theta_t, r_t) - L^\alpha(\theta_t^i, r_t)\right) & w_t^i > 0 \\
      \frac{1}{D}\left[L^\alpha(\theta_t, r_t) - L^\alpha(\theta_t^i, r_t))\right]_+ & w_t^i \leq 0 \\
    \end{cases}$.
    \State Update expert weight $w_{t+1}^i = \frac{1}{t - i + 1}\left( \sum_{j=i}^t g_j^i \right) \left(1 + \sum_{j=i}^t w_j^i g_j^i \right)$.
  \EndFor
\EndFor
\end{algorithmic}
\end{algorithm}

The Strongly Adaptive Online Conformal Prediction (SAOCP;
 Algorithm~\ref{algo-saocp} ) algorithm was proposed as an improvement
over the extant ACI algorithms in that it features stronger theoretical
guarantees. SAOCP works similarly to AgACI and FACI in that it maintains
a library of candidate online learning algorithms that generate
prediction intervals which are then aggregated using a meta-algorithm
(Bhatnagar et al. 2023). The candidate algorithm was chosen to be
SF-OGD, although any algorithm that features anytime regret guarantees
can be chosen. As opposed to AgACI and FACI, in which each candidate has
a different learning rate but is always able to contribute to the final
prediction intervals, here each candidate has the same learning rate but
only has positive weight over a specific interval of time. New candidate
algorithms are continually being spawned in order that, if the
distribution shifts rapidly, the newer candidates will be able to react
quickly and receive positive weight. Specifically, at each time point, a
new expert is instantiated which is active over a finite ``lifetime''.
Define the \emph{lifetime} of an expert instantiated at time \(t\) as \[
\begin{aligned}
  L(t) := g \cdot \max_{n \in \mathbb{Z}} \{ 2^n t \equiv 0 \mod 2^n \},
\end{aligned}
\] where \(g \in \mathbb{Z}^*\) is a \emph{lifetime multiplier}
parameter. The active experts are weighted according to their empirical
performance with respect to the pinball loss function. The authors show
that this construction results in intervals that have strong regret
guarantees.

\hypertarget{theoretical-guarantees-3}{%
\subsubsection{Theoretical Guarantees}\label{theoretical-guarantees-3}}

The theoretical results were established for SAOCP using the linear
interval constructor. The following bound for the strongly adaptive
regret holds for all subperiod lengths \(m \in \llbracket T \rrbracket\)
(Bhatnagar et al. 2023): \[
\begin{aligned}
  \mathrm{SAReg}(T, m) \leq 15 D \sqrt{m(\log T + 1)} \leq \tilde{\mathcal{O}}(D \sqrt m).
\end{aligned}
\] It should be emphasized that this regret bounds holds simultaneously
across all \(m\), as opposed to FACI, where a similar bound holds only
for a single \(m\). A bound on the coverage error of SAOCP has also been
established as: \[
\begin{aligned}
  |\mathrm{CovErr}(T)| \leq \mathcal{O}\left(\inf_\beta(T^{1/2 - \beta} + T^{\beta - 1} S_\beta(T))\right).
\end{aligned}
\] where \(S_{\beta}(T)\) is a technical measure of the smoothness of
the cumulative gradients and expert weights for each of the candidate
experts (Bhatnagar et al. 2023).

\hypertarget{tuning-parameters-4}{%
\subsubsection{Tuning Parameters}\label{tuning-parameters-4}}

\begin{figure}

{\centering \includegraphics{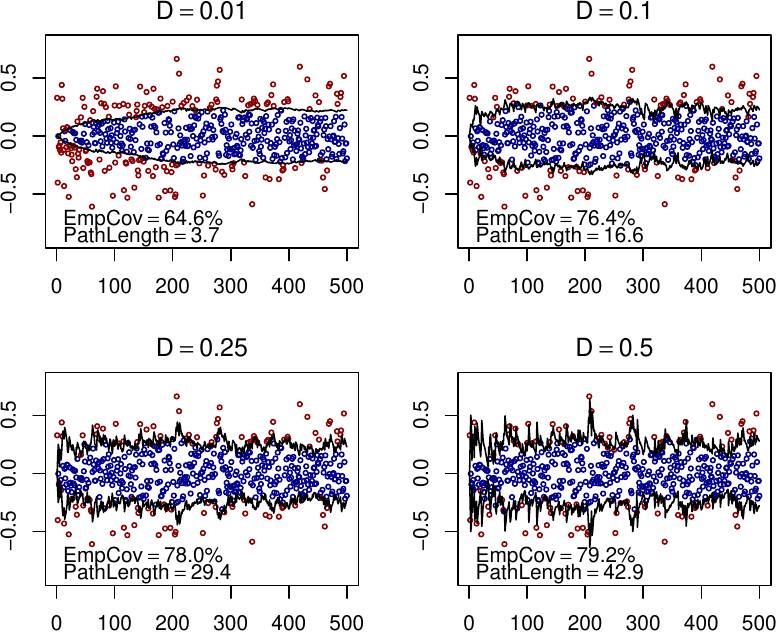}

}

\caption{\label{fig-saocp}Example 80\% prediction intervals generated by
the SAOCP algorithm with different values of the maximum radius
parameter \(D\). Blue and red points are observations that fell inside
and outside the prediction intervals, respectively.}

\end{figure}

The main tuning parameter for SAOCP is the learning rate \(\gamma\) of
the SF-OGD sub-algorithms, which we saw in the previous section has for
optimal choice \(\gamma = D / \sqrt{3}\). Values for \(D\) that are too
low lead to intervals that adapt slowly, and values that are too large
lead to jagged intervals. In their experiments, the authors select a
value for \(D\) by picking the maximum residual from a calibration set.
The second tuning parameter is the lifetime multiplier \(g\) which
controls the lifetime of each of the experts. We follow the original
paper in setting \(g = 8\). Figure~\ref{fig-saocp} illustrates the SAOCP
algorithm for choices of \(D \in \{0.01, 0.1, 0.25, 0.5 \}\).

\hypertarget{sec-simulations}{%
\section{Simulation Studies}\label{sec-simulations}}

We present two empirical studies in order to compare the performance of
the AgACI, FACI, SF-OGD, and SAOCP algorithms applied to simple
simulated datasets. The original ACI algorithm was not included as it is
not clear how to set the tuning rate \(\gamma\), which can have a large
effect on the resulting intervals. For both simulations we set the
targeted empirical coverage to \(\alpha = 0.8\), \(\alpha = 0.9\), and
\(\alpha = 0.95\). For each algorithm, we chose the interval constructor
that was used in its original presentation (see Table~\ref{tbl-aci}).

\hypertarget{time-series-with-arma-errors}{%
\subsection{Time series with ARMA
errors}\label{time-series-with-arma-errors}}

In this simulation we reproduce the setup described in Zaffran et al.
(2022) (itself based on that of Friedman, Grosse, and Stuetzle (1983)).
The time series values \(y_t\) for \(t \in \llbracket T \rrbracket\)
(\(T = 600\)) are simulated according to \[
\begin{aligned}
  y_t = 10\sin(\pi X_{t,1}X_{t,2}) + 20(X_{t,3} - 0.5)^2 + 10X_{t,4} + 5 X_{t,5} + 0X_{t,6} + \epsilon_t,
\end{aligned}
\] where \(X_{t,i}\), \(i = 1, \dots, 6\),
\(t \in \llbracket T \rrbracket\) are independently uniformly
distributed on \([0, 1]\) and the noise terms \(\epsilon_t\) are
generated according to an ARMA(1, 1) process: \[
\begin{aligned}
  \epsilon_t &= \psi \epsilon_{t-1} + \xi_t + \theta \xi_{t-1}, \\
  \xi_t &\sim N(0, \sigma^2).
\end{aligned}
\] We set \(\psi\) and \(\theta\) jointly to each value in
\(\{ 0.1, 0.8, 0.9, 0.95, 0.99 \}\) to simulate time series with
increasing temporal dependence. The innovation variance was set to
\(\sigma^2 = (1 - \psi^2) / (1 + 2\psi \xi + \xi^2)\) (to ensure that
the process has constant variance). For each setting, 25 simulated
datasets were generated.

To provide point predictions for the ACI algorithms, at each time
\(t \geq 200\) a random forest model was fitted to the previously
observed data using the \texttt{ranger} \texttt{R} package (Wright and
Ziegler 2017). The estimated model was then used to predict the
subsequent time point. The maximum radius \(D\) was estimated as the
maximum residual observed between time points \(t=200\) and \(t=249\).
The ACI models were then executed starting at time point \(t = 250\).
All metrics are based on time points \(t \geq 300\) to allow time for
the ACI methods to initialize.

\begin{Shaded}
\begin{Highlighting}[]
\NormalTok{simulate }\OtherTok{\textless{}{-}} \ControlFlowTok{function}\NormalTok{(seed, psi, xi, }\AttributeTok{N =} \FloatTok{1e3}\NormalTok{) \{}
  \FunctionTok{set.seed}\NormalTok{(seed)}
  
\NormalTok{  s }\OtherTok{\textless{}{-}} \DecValTok{10}
\NormalTok{  innov\_scale }\OtherTok{\textless{}{-}} \FunctionTok{sqrt}\NormalTok{(s }\SpecialCharTok{*}\NormalTok{ (}\DecValTok{1} \SpecialCharTok{{-}}\NormalTok{ psi}\SpecialCharTok{\^{}}\DecValTok{2}\NormalTok{) }\SpecialCharTok{/}\NormalTok{ (}\DecValTok{1} \SpecialCharTok{+} \DecValTok{2} \SpecialCharTok{*}\NormalTok{ psi }\SpecialCharTok{*}\NormalTok{ xi }\SpecialCharTok{+}\NormalTok{ xi}\SpecialCharTok{\^{}}\DecValTok{2}\NormalTok{))}
  
\NormalTok{  X }\OtherTok{\textless{}{-}} \FunctionTok{matrix}\NormalTok{(}\FunctionTok{runif}\NormalTok{(}\DecValTok{6} \SpecialCharTok{*}\NormalTok{ N), }\AttributeTok{ncol =} \DecValTok{6}\NormalTok{, }\AttributeTok{nrow =}\NormalTok{ N)}
  \FunctionTok{colnames}\NormalTok{(X) }\OtherTok{\textless{}{-}} \FunctionTok{paste0}\NormalTok{(}\StringTok{"X"}\NormalTok{, }\DecValTok{1}\SpecialCharTok{:}\DecValTok{6}\NormalTok{)}
  
\NormalTok{  epsilon }\OtherTok{\textless{}{-}} \FunctionTok{arima.sim}\NormalTok{(}\AttributeTok{n =}\NormalTok{ N, }\AttributeTok{model =} \FunctionTok{list}\NormalTok{(}\AttributeTok{ar =}\NormalTok{ psi, }\AttributeTok{ma =}\NormalTok{ xi), }\AttributeTok{sd =}\NormalTok{ innov\_scale)}
  
\NormalTok{  mu }\OtherTok{\textless{}{-}} \DecValTok{10} \SpecialCharTok{*} \FunctionTok{sin}\NormalTok{(pi }\SpecialCharTok{*}\NormalTok{ X[,}\DecValTok{1}\NormalTok{] }\SpecialCharTok{*}\NormalTok{ X[,}\DecValTok{2}\NormalTok{]) }\SpecialCharTok{+} \DecValTok{20} \SpecialCharTok{*}\NormalTok{ (X[,}\DecValTok{3}\NormalTok{] }\SpecialCharTok{{-}} \FloatTok{0.5}\NormalTok{)}\SpecialCharTok{\^{}}\DecValTok{2} \SpecialCharTok{+} \DecValTok{10} \SpecialCharTok{*}\NormalTok{ X[,}\DecValTok{4}\NormalTok{] }\SpecialCharTok{+} \DecValTok{5} \SpecialCharTok{*}\NormalTok{ X[,}\DecValTok{5}\NormalTok{]}
\NormalTok{  y }\OtherTok{\textless{}{-}}\NormalTok{ mu }\SpecialCharTok{+}\NormalTok{ epsilon}
  \FunctionTok{as\_tibble}\NormalTok{(X) }\SpecialCharTok{\%\textgreater{}\%} \FunctionTok{mutate}\NormalTok{(}\AttributeTok{y =}\NormalTok{ y)}
\NormalTok{\}}

\NormalTok{estimate\_model }\OtherTok{\textless{}{-}} \ControlFlowTok{function}\NormalTok{(data, }\AttributeTok{p =} \ConstantTok{NULL}\NormalTok{) \{}
  \ControlFlowTok{if}\NormalTok{(}\SpecialCharTok{!}\FunctionTok{is.null}\NormalTok{(p)) }\FunctionTok{p}\NormalTok{()}
\NormalTok{  preds }\OtherTok{\textless{}{-}} \FunctionTok{numeric}\NormalTok{(}\FunctionTok{nrow}\NormalTok{(data))}
  \ControlFlowTok{for}\NormalTok{(t }\ControlFlowTok{in} \DecValTok{200}\SpecialCharTok{:}\FunctionTok{nrow}\NormalTok{(data)) \{}
\NormalTok{    model }\OtherTok{\textless{}{-}}\NormalTok{ ranger}\SpecialCharTok{::}\FunctionTok{ranger}\NormalTok{(y }\SpecialCharTok{\textasciitilde{}}\NormalTok{ X1 }\SpecialCharTok{+}\NormalTok{ X2 }\SpecialCharTok{+}\NormalTok{ X3 }\SpecialCharTok{+}\NormalTok{ X4 }\SpecialCharTok{+}\NormalTok{ X5 }\SpecialCharTok{+}\NormalTok{ X6, }\AttributeTok{data =}\NormalTok{ data[}\DecValTok{1}\SpecialCharTok{:}\NormalTok{(t }\SpecialCharTok{{-}} \DecValTok{1}\NormalTok{),])}
\NormalTok{    preds[t] }\OtherTok{\textless{}{-}} \FunctionTok{predict}\NormalTok{(model, }\AttributeTok{data =}\NormalTok{ data[t, ])}\SpecialCharTok{$}\NormalTok{predictions}
\NormalTok{  \}}
\NormalTok{  preds}
\NormalTok{\}}

\NormalTok{metrics }\OtherTok{\textless{}{-}} \ControlFlowTok{function}\NormalTok{(fit) \{}
\NormalTok{  indices }\OtherTok{\textless{}{-}} \DecValTok{300}\SpecialCharTok{:}\FunctionTok{length}\NormalTok{(fit}\SpecialCharTok{$}\NormalTok{Y)}
  \FunctionTok{aci\_metrics}\NormalTok{(fit, indices)}
\NormalTok{\}}

\NormalTok{fit }\OtherTok{\textless{}{-}} \ControlFlowTok{function}\NormalTok{(data, preds, method, alpha, }\AttributeTok{p =} \ConstantTok{NULL}\NormalTok{) \{}
  \ControlFlowTok{if}\NormalTok{(}\SpecialCharTok{!}\FunctionTok{is.null}\NormalTok{(p)) }\FunctionTok{p}\NormalTok{()}
  
\NormalTok{  D }\OtherTok{\textless{}{-}} \FunctionTok{max}\NormalTok{(}\FunctionTok{abs}\NormalTok{(data}\SpecialCharTok{$}\NormalTok{y }\SpecialCharTok{{-}}\NormalTok{ preds)[}\DecValTok{200}\SpecialCharTok{:}\DecValTok{249}\NormalTok{])}
\NormalTok{  gamma }\OtherTok{\textless{}{-}}\NormalTok{ D }\SpecialCharTok{/} \FunctionTok{sqrt}\NormalTok{(}\DecValTok{3}\NormalTok{)}
  
\NormalTok{  interval\_constructor }\OtherTok{=} \FunctionTok{case\_when}\NormalTok{(}
\NormalTok{    method }\SpecialCharTok{==} \StringTok{"AgACI"} \SpecialCharTok{\textasciitilde{}} \StringTok{"conformal"}\NormalTok{,}
\NormalTok{    method }\SpecialCharTok{==} \StringTok{"FACI"} \SpecialCharTok{\textasciitilde{}} \StringTok{"conformal"}\NormalTok{,}
\NormalTok{    method }\SpecialCharTok{==} \StringTok{"SF{-}OGD"} \SpecialCharTok{\textasciitilde{}} \StringTok{"linear"}\NormalTok{,}
\NormalTok{    method }\SpecialCharTok{==} \StringTok{"SAOCP"} \SpecialCharTok{\textasciitilde{}} \StringTok{"linear"}
\NormalTok{  )}
  
  \ControlFlowTok{if}\NormalTok{(interval\_constructor }\SpecialCharTok{==} \StringTok{"linear"}\NormalTok{) \{}
\NormalTok{    gamma\_grid }\OtherTok{=} \FunctionTok{seq}\NormalTok{(}\FloatTok{0.1}\NormalTok{, }\DecValTok{1}\NormalTok{, }\FloatTok{0.1}\NormalTok{)}
\NormalTok{  \}}
  \ControlFlowTok{else}\NormalTok{ \{}
\NormalTok{    gamma\_grid }\OtherTok{\textless{}{-}} \FunctionTok{c}\NormalTok{(}\FloatTok{0.001}\NormalTok{, }\FloatTok{0.002}\NormalTok{, }\FloatTok{0.004}\NormalTok{, }\FloatTok{0.008}\NormalTok{, }\FloatTok{0.016}\NormalTok{, }\FloatTok{0.032}\NormalTok{, }\FloatTok{0.064}\NormalTok{, }\FloatTok{0.128}\NormalTok{)}
\NormalTok{  \}}
  
\NormalTok{  parameters }\OtherTok{\textless{}{-}} \FunctionTok{list}\NormalTok{(}
    \AttributeTok{interval\_constructor =}\NormalTok{ interval\_constructor, }
    \AttributeTok{D =}\NormalTok{ D, }
    \AttributeTok{gamma =}\NormalTok{ gamma, }
    \AttributeTok{gamma\_grid =}\NormalTok{ gamma\_grid}
\NormalTok{  )}
  
  \FunctionTok{aci}\NormalTok{(}
\NormalTok{    data}\SpecialCharTok{$}\NormalTok{y[}\DecValTok{250}\SpecialCharTok{:}\FunctionTok{nrow}\NormalTok{(data)], }
\NormalTok{    preds[}\DecValTok{250}\SpecialCharTok{:}\FunctionTok{nrow}\NormalTok{(data)], }
    \AttributeTok{method =}\NormalTok{ method, }
    \AttributeTok{alpha =}\NormalTok{ alpha, }
    \AttributeTok{parameters =}\NormalTok{ parameters}
\NormalTok{  )}
\NormalTok{\}}

\NormalTok{N\_sims }\OtherTok{\textless{}{-}} \DecValTok{25}
\NormalTok{simulation\_data }\OtherTok{\textless{}{-}} \FunctionTok{expand\_grid}\NormalTok{(}
  \AttributeTok{index =} \DecValTok{1}\SpecialCharTok{:}\NormalTok{N\_sims,}
  \AttributeTok{param =}  \FunctionTok{c}\NormalTok{(}\FloatTok{0.1}\NormalTok{, }\FloatTok{0.8}\NormalTok{, }\FloatTok{0.9}\NormalTok{, }\FloatTok{0.95}\NormalTok{, }\FloatTok{0.99}\NormalTok{),}
  \AttributeTok{N =} \DecValTok{600}
\NormalTok{) }\SpecialCharTok{\%\textgreater{}\%}
  \FunctionTok{mutate}\NormalTok{(}\AttributeTok{psi =}\NormalTok{ param, }\AttributeTok{xi =}\NormalTok{ param)}

\CommentTok{\# For each simulated dataset, fit multiple ACI methods}
\NormalTok{simulation\_study\_setup }\OtherTok{\textless{}{-}} \FunctionTok{expand\_grid}\NormalTok{(}
  \AttributeTok{alpha =} \FunctionTok{c}\NormalTok{(}\FloatTok{0.8}\NormalTok{, }\FloatTok{0.9}\NormalTok{, }\FloatTok{0.95}\NormalTok{),}
  \AttributeTok{method =} \FunctionTok{c}\NormalTok{(}\StringTok{"AgACI"}\NormalTok{, }\StringTok{"SF{-}OGD"}\NormalTok{, }\StringTok{"SAOCP"}\NormalTok{, }\StringTok{"FACI"}\NormalTok{)}
\NormalTok{)}

\NormalTok{simulation\_study1 }\OtherTok{\textless{}{-}} \FunctionTok{run\_simulation\_study1}\NormalTok{(}
\NormalTok{  simulation\_data,}
\NormalTok{  simulation\_study\_setup,}
\NormalTok{  estimate\_model,}
\NormalTok{  fit,}
  \AttributeTok{workers =} \DecValTok{8}
\NormalTok{)}
\end{Highlighting}
\end{Shaded}

\begin{figure}[H]

{\centering \includegraphics{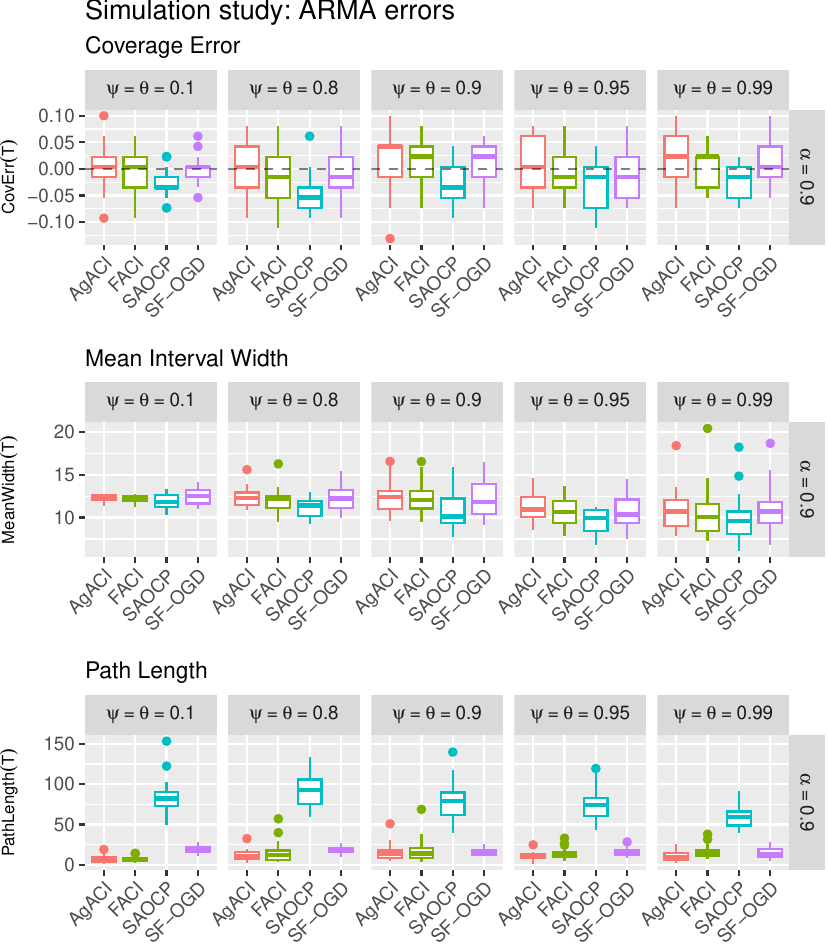}

}

\caption{\label{fig-simulation-one-results}Coverage errors, mean
interval widths, and path lengths for the first simulation study with
target coverage \(\alpha = 0.9\).}

\end{figure}

\begin{figure}[H]

{\centering \includegraphics{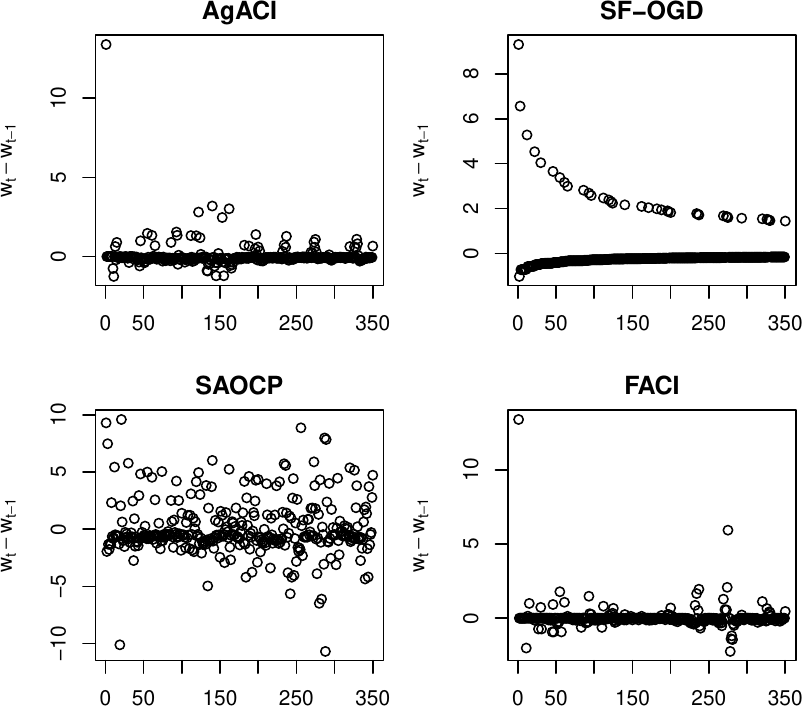}

}

\caption{\label{fig-simulation-one-example}Difference in successive
interval widths (\(w_t - w_{t-1}\)) from an illustrative simulation from
the first simulation study.}

\end{figure}

The coverage errors, mean interval widths, and path lengths of each of
the algorithms for \(\alpha = 0.9\) are shown in
Figure~\ref{fig-simulation-one-results} (results for
\(\alpha \in \{ 0.8, 0.95 \}\) were similar and are available in the
appendix). All methods achieved near optimal empirical coverage,
although SAOCP tended to slightly undercover. The mean interval widths
were similar across methods, although again SAOCP had slightly shorter
intervals (as could be expected given its tendency to undercover). The
path length of SAOCP was larger than any of the other methods. To
investigate why, Figure~\ref{fig-simulation-one-example} plots
\(w_t - w_{t-1}\), the difference in interval width between times
\(t-1\) and \(t\), for each method in one of the simulations. The
interval widths for AgACI and FACI change slowly relative to those for
SF-OGD and SAOCP. For SAOCP, we can see the interval widths have larger
fluctuations than for the other methods, explaining its higher path
width.

\begin{Shaded}
\begin{Highlighting}[]
\FunctionTok{simulation\_one\_plot}\NormalTok{(simulation\_study1}\SpecialCharTok{$}\NormalTok{results }\SpecialCharTok{\%\textgreater{}\%} \FunctionTok{filter}\NormalTok{(alpha }\SpecialCharTok{==} \FloatTok{0.9}\NormalTok{))}
\end{Highlighting}
\end{Shaded}

\begin{Shaded}
\begin{Highlighting}[]
\NormalTok{fits }\OtherTok{\textless{}{-}}\NormalTok{ simulation\_study1}\SpecialCharTok{$}\NormalTok{example\_fits}

\FunctionTok{par}\NormalTok{(}\AttributeTok{mfrow =} \FunctionTok{c}\NormalTok{(}\DecValTok{2}\NormalTok{, }\DecValTok{2}\NormalTok{), }\AttributeTok{mar =} \FunctionTok{c}\NormalTok{(}\DecValTok{3}\NormalTok{, }\DecValTok{4}\NormalTok{, }\DecValTok{2}\NormalTok{, }\DecValTok{1}\NormalTok{))}
\ControlFlowTok{for}\NormalTok{(i }\ControlFlowTok{in} \DecValTok{1}\SpecialCharTok{:}\DecValTok{4}\NormalTok{) \{}
  \FunctionTok{plot}\NormalTok{(}
    \FunctionTok{diff}\NormalTok{(fits}\SpecialCharTok{$}\NormalTok{fit[[i]]}\SpecialCharTok{$}\NormalTok{intervals[,}\DecValTok{2}\NormalTok{] }\SpecialCharTok{{-}}\NormalTok{ fits}\SpecialCharTok{$}\NormalTok{fit[[i]]}\SpecialCharTok{$}\NormalTok{intervals[,}\DecValTok{1}\NormalTok{]), }
    \AttributeTok{main =}\NormalTok{ fits}\SpecialCharTok{$}\NormalTok{method[[i]], }
    \AttributeTok{xlab =} \StringTok{"T"}\NormalTok{, }
    \AttributeTok{ylab =} \FunctionTok{expression}\NormalTok{(w[t] }\SpecialCharTok{{-}}\NormalTok{ w[t }\SpecialCharTok{{-}} \DecValTok{1}\NormalTok{]))}
\NormalTok{\}}
\FunctionTok{par}\NormalTok{(}\AttributeTok{mfrow =} \FunctionTok{c}\NormalTok{(}\DecValTok{1}\NormalTok{, }\DecValTok{1}\NormalTok{), }\AttributeTok{mar =} \FunctionTok{c}\NormalTok{(}\FloatTok{5.1}\NormalTok{, }\FloatTok{4.1}\NormalTok{, }\FloatTok{4.1}\NormalTok{, }\FloatTok{2.1}\NormalTok{))}
\end{Highlighting}
\end{Shaded}

\hypertarget{distribution-shift}{%
\subsection{Distribution shift}\label{distribution-shift}}

This simulation study features time series with distribution shifts. The
setup is quite simple in order to probe the basic performance of the
methods in response to distribution shift. As a baseline, we simulate
time series of independent data with \[
\begin{aligned}
  y_t &\sim N(0, \sigma_t^2), \\
  \sigma_t &= 0.2,
\end{aligned}
\] for all \(t \in \llbracket T \rrbracket\) (\(T = 500\)). In the
second type of time series, the observations are still independent but
their variance increases halfway through the time series: \[
\begin{aligned}
y_t &\sim N(0, \sigma_t^2), \\
\sigma_t &= 0.2 + 0.5 \mathbb{I}[t > 250].
\end{aligned}
\] In each case, the ACI algorithms are provided with the unbiased
predictions \(\hat{\mu}_t = 0\), \(t \in \llbracket T \rrbracket\).
Fifty simulated datasets were generated for each type of time series.

\begin{Shaded}
\begin{Highlighting}[]
\NormalTok{simulate }\OtherTok{\textless{}{-}} \ControlFlowTok{function}\NormalTok{(seed, }\AttributeTok{distribution\_shift =} \DecValTok{0}\NormalTok{, }\AttributeTok{N =} \FloatTok{1e3}\NormalTok{, }\AttributeTok{sigma =} \FloatTok{0.2}\NormalTok{) \{}
  \FunctionTok{set.seed}\NormalTok{(seed)}
\NormalTok{  mu }\OtherTok{\textless{}{-}} \FunctionTok{rep}\NormalTok{(}\DecValTok{0}\NormalTok{, N)}
\NormalTok{  shift }\OtherTok{\textless{}{-}} \DecValTok{1}\SpecialCharTok{:}\NormalTok{N }\SpecialCharTok{\textgreater{}}\NormalTok{ (N }\SpecialCharTok{/} \DecValTok{2}\NormalTok{)}
\NormalTok{  muhat }\OtherTok{\textless{}{-}}\NormalTok{ mu}
\NormalTok{  y }\OtherTok{\textless{}{-}} \FunctionTok{rnorm}\NormalTok{(}\AttributeTok{n =} \FunctionTok{length}\NormalTok{(mu), }\AttributeTok{mean =}\NormalTok{ mu, }\AttributeTok{sd =}\NormalTok{ sigma }\SpecialCharTok{+} \FunctionTok{ifelse}\NormalTok{(shift, distribution\_shift, }\DecValTok{0}\NormalTok{))}
  
  \FunctionTok{tibble}\NormalTok{(}\AttributeTok{y =}\NormalTok{ y, }\AttributeTok{muhat =}\NormalTok{ muhat)}
\NormalTok{\}}

\NormalTok{metrics }\OtherTok{\textless{}{-}} \ControlFlowTok{function}\NormalTok{(fit) \{}
\NormalTok{  N }\OtherTok{\textless{}{-}} \FunctionTok{length}\NormalTok{(fit}\SpecialCharTok{$}\NormalTok{Y)}
\NormalTok{  indices }\OtherTok{\textless{}{-}} \FunctionTok{which}\NormalTok{(}\DecValTok{1}\SpecialCharTok{:}\NormalTok{N }\SpecialCharTok{\textgreater{}} \DecValTok{50}\NormalTok{)}
  \FunctionTok{aci\_metrics}\NormalTok{(fit, indices)}
\NormalTok{\}}

\NormalTok{fit }\OtherTok{\textless{}{-}} \ControlFlowTok{function}\NormalTok{(data, method, alpha, }\AttributeTok{p =} \ConstantTok{NULL}\NormalTok{) \{}
  \ControlFlowTok{if}\NormalTok{(}\SpecialCharTok{!}\FunctionTok{is.null}\NormalTok{(p)) }\FunctionTok{p}\NormalTok{()}
  
\NormalTok{  interval\_constructor }\OtherTok{=} \FunctionTok{case\_when}\NormalTok{(}
\NormalTok{    method }\SpecialCharTok{==} \StringTok{"AgACI"} \SpecialCharTok{\textasciitilde{}} \StringTok{"conformal"}\NormalTok{,}
\NormalTok{    method }\SpecialCharTok{==} \StringTok{"FACI"} \SpecialCharTok{\textasciitilde{}} \StringTok{"conformal"}\NormalTok{,}
\NormalTok{    method }\SpecialCharTok{==} \StringTok{"SF{-}OGD"} \SpecialCharTok{\textasciitilde{}} \StringTok{"linear"}\NormalTok{,}
\NormalTok{    method }\SpecialCharTok{==} \StringTok{"SAOCP"} \SpecialCharTok{\textasciitilde{}} \StringTok{"linear"}
\NormalTok{  )}
  
  \ControlFlowTok{if}\NormalTok{(interval\_constructor }\SpecialCharTok{==} \StringTok{"linear"}\NormalTok{) \{}
\NormalTok{    D }\OtherTok{\textless{}{-}} \FunctionTok{max}\NormalTok{(}\FunctionTok{abs}\NormalTok{(data}\SpecialCharTok{$}\NormalTok{y }\SpecialCharTok{{-}}\NormalTok{ data}\SpecialCharTok{$}\NormalTok{muhat)[}\DecValTok{1}\SpecialCharTok{:}\DecValTok{50}\NormalTok{])}
\NormalTok{  \}}
  \ControlFlowTok{else}\NormalTok{ \{}
\NormalTok{    D }\OtherTok{\textless{}{-}} \DecValTok{1}
\NormalTok{  \}}
  
\NormalTok{  gamma }\OtherTok{\textless{}{-}}\NormalTok{ D }\SpecialCharTok{/} \FunctionTok{sqrt}\NormalTok{(}\DecValTok{3}\NormalTok{)}
  
  \ControlFlowTok{if}\NormalTok{(interval\_constructor }\SpecialCharTok{==} \StringTok{"linear"}\NormalTok{) \{}
\NormalTok{    gamma\_grid }\OtherTok{\textless{}{-}} \FunctionTok{seq}\NormalTok{(}\FloatTok{0.1}\NormalTok{, }\DecValTok{2}\NormalTok{, }\FloatTok{0.1}\NormalTok{)}
\NormalTok{  \}}
  \ControlFlowTok{else}\NormalTok{ \{}
\NormalTok{    gamma\_grid }\OtherTok{\textless{}{-}} \FunctionTok{c}\NormalTok{(}\FloatTok{0.001}\NormalTok{, }\FloatTok{0.002}\NormalTok{, }\FloatTok{0.004}\NormalTok{, }\FloatTok{0.008}\NormalTok{, }\FloatTok{0.016}\NormalTok{, }\FloatTok{0.032}\NormalTok{, }\FloatTok{0.064}\NormalTok{, }\FloatTok{0.128}\NormalTok{)}
\NormalTok{  \}}
  
\NormalTok{  parameters }\OtherTok{\textless{}{-}} \FunctionTok{list}\NormalTok{(}
    \AttributeTok{interval\_constructor =}\NormalTok{ interval\_constructor, }
    \AttributeTok{D =}\NormalTok{ D, }
    \AttributeTok{gamma =}\NormalTok{ gamma, }
    \AttributeTok{gamma\_grid =}\NormalTok{ gamma\_grid}
\NormalTok{  )}
  
  \FunctionTok{aci}\NormalTok{(data}\SpecialCharTok{$}\NormalTok{y, data}\SpecialCharTok{$}\NormalTok{muhat, }\AttributeTok{method =}\NormalTok{ method, }\AttributeTok{alpha =}\NormalTok{ alpha, }\AttributeTok{parameters =}\NormalTok{ parameters)}
\NormalTok{\}}

\NormalTok{N\_sims }\OtherTok{\textless{}{-}} \FloatTok{5e1}
\NormalTok{simulation\_study\_setup2 }\OtherTok{\textless{}{-}} \FunctionTok{expand\_grid}\NormalTok{(}
  \AttributeTok{index =} \DecValTok{1}\SpecialCharTok{:}\NormalTok{N\_sims,}
  \AttributeTok{distribution\_shift =} \FunctionTok{c}\NormalTok{(}\DecValTok{0}\NormalTok{, }\FloatTok{0.5}\NormalTok{),}
  \AttributeTok{alpha =} \FunctionTok{c}\NormalTok{(}\FloatTok{0.8}\NormalTok{, }\FloatTok{0.9}\NormalTok{, }\FloatTok{0.95}\NormalTok{),}
  \AttributeTok{N =} \DecValTok{500}\NormalTok{,}
  \AttributeTok{method =} \FunctionTok{c}\NormalTok{(}\StringTok{"AgACI"}\NormalTok{, }\StringTok{"SF{-}OGD"}\NormalTok{, }\StringTok{"SAOCP"}\NormalTok{, }\StringTok{"FACI"}\NormalTok{),}
\NormalTok{) }\SpecialCharTok{\%\textgreater{}\%}
  \FunctionTok{mutate}\NormalTok{(}\AttributeTok{data =} \FunctionTok{pmap}\NormalTok{(}\FunctionTok{list}\NormalTok{(index, distribution\_shift, N), simulate))}

\NormalTok{simulation\_study2 }\OtherTok{\textless{}{-}} \FunctionTok{run\_simulation\_study2}\NormalTok{(simulation\_study\_setup2, fit, }\AttributeTok{workers =} \DecValTok{8}\NormalTok{)}
\end{Highlighting}
\end{Shaded}

\begin{figure}[H]

{\centering \includegraphics{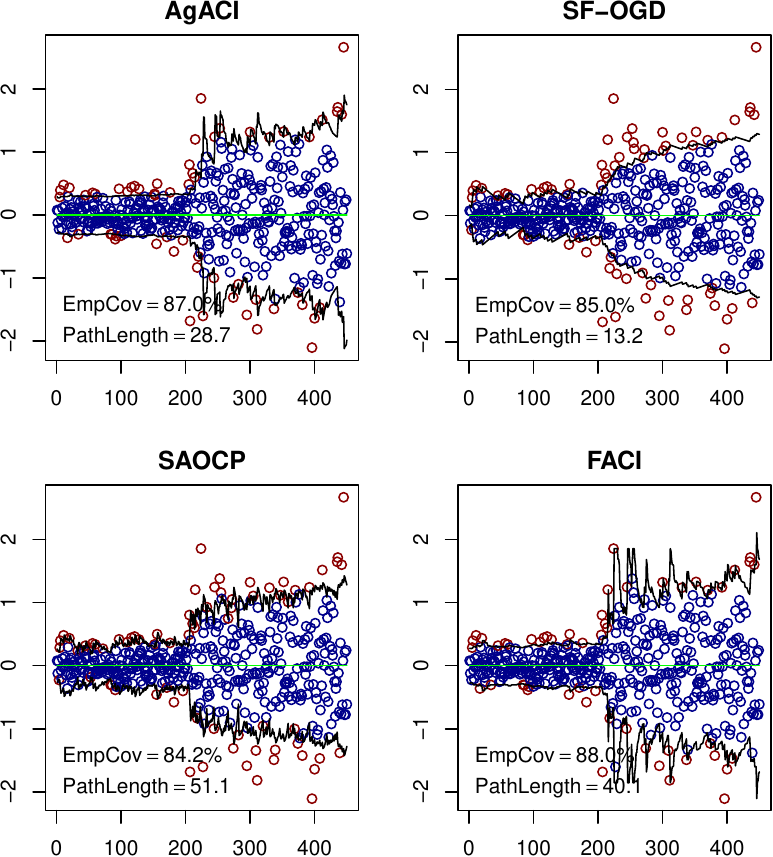}

}

\caption{\label{fig-simulation-two-example}Example prediction intervals
(target coverage \(\alpha = 0.9\)) from the second simulation study of
time series with distributional shift, in which the shift occurs at time
250. Blue and red points are observations that fell inside and outside
the prediction intervals, respectively.}

\end{figure}

The coverage error, mean path length, and mean interval widths of the
algorithms are summarized in Figure~\ref{fig-simulation-two-joint} (an
alternative plot is included in the appendix as
Figure~\ref{fig-simulation-two-results}). The coverage error of all the
algorithms is near the desired value in the absence of distribution
shift. On the contrary, all of the algorithms except AgACI and FACI
undercover when there is distributional shift. SAOCP tends to have
higher average path lengths than the other methods. An illustrative
example of prediction intervals generated by each method for one of the
simulated time series with distribution shift is shown in
Figure~\ref{fig-simulation-two-example}. The SAOCP prediction intervals
in the example before the distribution shift are more jagged than those
produced by the other methods, which illustrates why SAOCP may have
longer path lengths.

\begin{figure}

{\centering \includegraphics{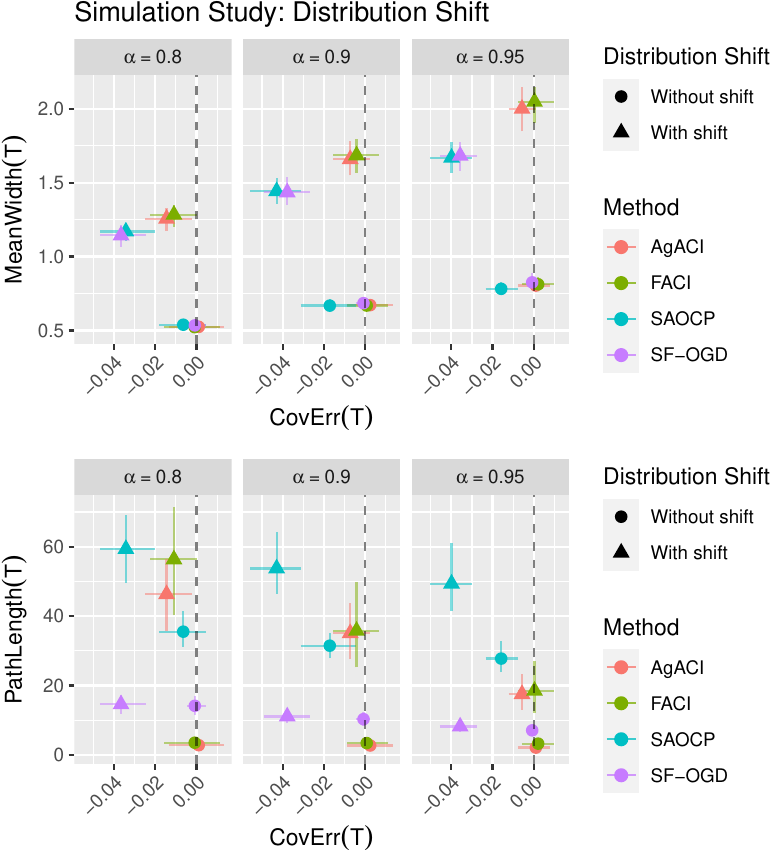}

}

\caption{\label{fig-simulation-two-joint}Mean interval width vs coverage
error (top) and Mean Path Length vs.~coverage error (bottom) for the
second simulation study. The error bars represent the 10\% to 90\%
quantiles of the metrics over the simulation datasets.}

\end{figure}

\begin{Shaded}
\begin{Highlighting}[]
\NormalTok{fits }\OtherTok{\textless{}{-}}\NormalTok{ simulation\_study2}\SpecialCharTok{$}\NormalTok{example\_fits}

\NormalTok{coverage    }\OtherTok{\textless{}{-}} \FunctionTok{format\_coverage}\NormalTok{(}\FunctionTok{extract\_metric}\NormalTok{(fits}\SpecialCharTok{$}\NormalTok{fit, }\StringTok{"coverage"}\NormalTok{))}
\NormalTok{path\_length }\OtherTok{\textless{}{-}} \FunctionTok{format\_path\_length}\NormalTok{(}\FunctionTok{extract\_metric}\NormalTok{(fits}\SpecialCharTok{$}\NormalTok{fit, }\StringTok{"path\_length"}\NormalTok{))}

\FunctionTok{par}\NormalTok{(}\AttributeTok{mfrow =} \FunctionTok{c}\NormalTok{(}\DecValTok{2}\NormalTok{, }\DecValTok{2}\NormalTok{), }\AttributeTok{mar =} \FunctionTok{c}\NormalTok{(}\DecValTok{3}\NormalTok{, }\DecValTok{3}\NormalTok{, }\DecValTok{2}\NormalTok{, }\DecValTok{1}\NormalTok{))}
\ControlFlowTok{for}\NormalTok{(i }\ControlFlowTok{in} \DecValTok{1}\SpecialCharTok{:}\DecValTok{4}\NormalTok{) \{}
  \FunctionTok{plot}\NormalTok{(fits}\SpecialCharTok{$}\NormalTok{fit[[i]], }\AttributeTok{legend =} \ConstantTok{FALSE}\NormalTok{, }\AttributeTok{main =}\NormalTok{ fits}\SpecialCharTok{$}\NormalTok{method[[i]], }\AttributeTok{index =} \DecValTok{51}\SpecialCharTok{:}\DecValTok{500}\NormalTok{)}
  \FunctionTok{text}\NormalTok{(}\AttributeTok{x =} \SpecialCharTok{{-}}\DecValTok{10}\NormalTok{, }\AttributeTok{y =} \SpecialCharTok{{-}}\FloatTok{1.5}\NormalTok{, }\AttributeTok{labels =} \FunctionTok{bquote}\NormalTok{(EmpCov }\SpecialCharTok{==}\NormalTok{ .(coverage[[i]]) ), }\AttributeTok{pos =} \DecValTok{4}\NormalTok{)}
  \FunctionTok{text}\NormalTok{(}\AttributeTok{x =} \SpecialCharTok{{-}}\DecValTok{10}\NormalTok{, }\AttributeTok{y =} \SpecialCharTok{{-}}\DecValTok{2}\NormalTok{, }\AttributeTok{labels =} \FunctionTok{bquote}\NormalTok{(PathLength }\SpecialCharTok{==}\NormalTok{ .(path\_length[[i]]) ), }\AttributeTok{pos =} \DecValTok{4}\NormalTok{)}
\NormalTok{\}}
\FunctionTok{par}\NormalTok{(}\AttributeTok{mfrow =} \FunctionTok{c}\NormalTok{(}\DecValTok{1}\NormalTok{, }\DecValTok{1}\NormalTok{), }\AttributeTok{mar =} \FunctionTok{c}\NormalTok{(}\FloatTok{5.1}\NormalTok{, }\FloatTok{4.1}\NormalTok{, }\FloatTok{4.1}\NormalTok{, }\FloatTok{2.1}\NormalTok{))}
\end{Highlighting}
\end{Shaded}

\hypertarget{sec-case-study}{%
\section{Case Study: Influenza Forecasting}\label{sec-case-study}}

Influenza is a highly infectious disease that is estimated to infect
approximately one billion individuals each year around the world
(Krammer et al. 2018). Influenza incidence in temperate climates tends
to follow a seasonal pattern, with the highest number of infections
during what is commonly referred to as the \textit{flu season} (Lofgren
et al. 2007). Accurate forecasting of influenza is of significant
interest to aid in public health planning and resource allocation. To
investigate the accuracy of influenza forecasts, the US Centers for
Disease Control (CDC) initiated a challenge, referred to as FluSight, in
which teams from multiple institutions submitted weekly forecasts of
influenza incidence (Biggerstaff et al. 2016). Reich et al. (2019)
evaluated the accuracy of the forecasts over seven flu seasons from 2010
to 2017. As a case study, we investigate the use of ACI algorithms to
augment the FluSight forecasts with prediction intervals.

The FluSight challenge collected forecasts for multiple prediction
targets. For this case study, we focus on national (US) one-week ahead
forecasts of weighted influenza-like illness (wILI), which is a
population-weighted percentage of doctors visits where patients
presented with influenza-like symptoms (Biggerstaff et al. 2016). The
FluSight dataset, which is publicy available, include forecasts derived
from 21 different forecasting models, from both mechanistic and
statistical viewpoints (Flusight Network 2020; Tushar et al. 2018,
2019). For our purposes, we treat the way the forecasts were produced as
a black box.

Formally, let \(y_{t}\), \(t \in \llbracket T \rrbracket\) be the
observed national wILI at time \(t\), and let \(\hat{\mu}_{j,t}\),
\(j \in \llbracket J \rrbracket\), be the one-week ahead forecast of the
wILI from model \(j\) at time \(t\). Two of the original 21 forecasting
methods were excluded from this case study due to poor predictive
performance (\texttt{Delphi\_Uniform} and \texttt{CUBMA}). In addition,
six methods had identical forecasts (\texttt{CU\_EAKFC\_SIRS},
\texttt{CU\_EKF\_SEIRS}, \texttt{CU\_EKF\_SIRS},
\texttt{CU\_RHF\_SEIRS}, \texttt{CU\_RHF\_SIRS}), and therefore we only
included one (\texttt{CU\_EAKFC\_SEIRS}) in the analysis. The ACI
methods were then applied to the log-observations and log-predictions,
where the log-transformation was used to constrain the final prediction
intervals to be positive. The first flu season (2010-2011) was used as a
warm-up for each ACI method, and we report the empirical performance of
the prediction intervals for the subsequent seasons (six seasons from
2012-2013 to 2016-2017). The ACI algorithms target prediction intervals
with coverage of \(\alpha = 0.8\), \(\alpha = 0.9\), and
\(\alpha = 0.95\). As in the simulation study, we used the interval
constructor corresponding to the original presentaiton of each algorithm
(see Table~\ref{tbl-aci}).

\begin{Shaded}
\begin{Highlighting}[]
\NormalTok{url }\OtherTok{\textless{}{-}} \StringTok{"https://raw.githubusercontent.com/FluSightNetwork/cdc{-}flusight{-}ensemble/master/scores/point\_ests.csv"}
\NormalTok{raw\_data }\OtherTok{\textless{}{-}} \FunctionTok{read\_csv}\NormalTok{(url, }\AttributeTok{show\_col\_types =} \ConstantTok{FALSE}\NormalTok{)}

\NormalTok{fit }\OtherTok{\textless{}{-}} \ControlFlowTok{function}\NormalTok{(data, method, alpha) \{}
\NormalTok{  first\_season }\OtherTok{\textless{}{-}}\NormalTok{ data}\SpecialCharTok{$}\NormalTok{Season }\SpecialCharTok{==} \StringTok{"2010/2011"}
\NormalTok{  D }\OtherTok{\textless{}{-}} \FunctionTok{max}\NormalTok{(}\FunctionTok{abs}\NormalTok{(data}\SpecialCharTok{$}\NormalTok{obs\_value }\SpecialCharTok{{-}}\NormalTok{ data}\SpecialCharTok{$}\NormalTok{Value)[first\_season])}
  
\NormalTok{  interval\_constructor }\OtherTok{=} \FunctionTok{case\_when}\NormalTok{(}
\NormalTok{    method }\SpecialCharTok{==} \StringTok{"AgACI"} \SpecialCharTok{\textasciitilde{}} \StringTok{"conformal"}\NormalTok{,}
\NormalTok{    method }\SpecialCharTok{==} \StringTok{"FACI"} \SpecialCharTok{\textasciitilde{}} \StringTok{"conformal"}\NormalTok{,}
\NormalTok{    method }\SpecialCharTok{==} \StringTok{"SF{-}OGD"} \SpecialCharTok{\textasciitilde{}} \StringTok{"linear"}\NormalTok{,}
\NormalTok{    method }\SpecialCharTok{==} \StringTok{"SAOCP"} \SpecialCharTok{\textasciitilde{}} \StringTok{"linear"}
\NormalTok{  )}
  
\NormalTok{  gamma }\OtherTok{\textless{}{-}}\NormalTok{ D }\SpecialCharTok{/} \FunctionTok{sqrt}\NormalTok{(}\DecValTok{3}\NormalTok{)}
  
  \ControlFlowTok{if}\NormalTok{(interval\_constructor }\SpecialCharTok{==} \StringTok{"linear"}\NormalTok{) \{}
\NormalTok{    gamma\_grid }\OtherTok{=} \FunctionTok{seq}\NormalTok{(}\FloatTok{0.1}\NormalTok{, }\DecValTok{1}\NormalTok{, }\FloatTok{0.1}\NormalTok{)}
\NormalTok{  \}}
  \ControlFlowTok{else}\NormalTok{ \{}
\NormalTok{    gamma\_grid }\OtherTok{\textless{}{-}} \FunctionTok{c}\NormalTok{(}\FloatTok{0.001}\NormalTok{, }\FloatTok{0.002}\NormalTok{, }\FloatTok{0.004}\NormalTok{, }\FloatTok{0.008}\NormalTok{, }\FloatTok{0.016}\NormalTok{, }\FloatTok{0.032}\NormalTok{, }\FloatTok{0.064}\NormalTok{, }\FloatTok{0.128}\NormalTok{)}
\NormalTok{  \}}
  
\NormalTok{  parameters }\OtherTok{\textless{}{-}} \FunctionTok{list}\NormalTok{(}
    \AttributeTok{interval\_constructor =}\NormalTok{ interval\_constructor,}
    \AttributeTok{D =}\NormalTok{ D, }
    \AttributeTok{gamma =}\NormalTok{ gamma, }
    \AttributeTok{gamma\_grid =}\NormalTok{ gamma\_grid}
\NormalTok{  )}
  
  \FunctionTok{aci}\NormalTok{(}
    \AttributeTok{Y =} \FunctionTok{log}\NormalTok{(data}\SpecialCharTok{$}\NormalTok{obs\_value), }
    \AttributeTok{predictions =} \FunctionTok{log}\NormalTok{(data}\SpecialCharTok{$}\NormalTok{Value), }
    \AttributeTok{method =}\NormalTok{ method, }
    \AttributeTok{parameters =}\NormalTok{ parameters, }
    \AttributeTok{alpha =}\NormalTok{ alpha}
\NormalTok{  )}
\NormalTok{\}}

\NormalTok{metrics }\OtherTok{\textless{}{-}} \ControlFlowTok{function}\NormalTok{(data, fit) \{}
  \FunctionTok{aci\_metrics}\NormalTok{(fit, }\AttributeTok{indices =} \FunctionTok{which}\NormalTok{(data}\SpecialCharTok{$}\NormalTok{Season }\SpecialCharTok{!=} \StringTok{"2010/2011"}\NormalTok{))}
\NormalTok{\}}

\NormalTok{analysis\_data }\OtherTok{\textless{}{-}}\NormalTok{ raw\_data }\SpecialCharTok{\%\textgreater{}\%}
  \FunctionTok{filter}\NormalTok{(}
\NormalTok{    Target }\SpecialCharTok{==} \StringTok{"1 wk ahead"}\NormalTok{, }
\NormalTok{    Location }\SpecialCharTok{==} \StringTok{"US National"}\NormalTok{, }
    \SpecialCharTok{!}\NormalTok{(model\_name }\SpecialCharTok{\%in\%} \FunctionTok{c}\NormalTok{(}\StringTok{"Delphi\_Uniform"}\NormalTok{, }\StringTok{"CUBMA"}\NormalTok{, }\StringTok{"CU\_EAKFC\_SIRS"}\NormalTok{, }\StringTok{"CU\_EKF\_SEIRS"}\NormalTok{, }\StringTok{"CU\_EKF\_SIRS"}\NormalTok{, }\StringTok{"CU\_RHF\_SEIRS"}\NormalTok{, }\StringTok{"CU\_RHF\_SIRS"}\NormalTok{))}
\NormalTok{  ) }\SpecialCharTok{\%\textgreater{}\%}
  \FunctionTok{arrange}\NormalTok{(Year, Model.Week) }\SpecialCharTok{\%\textgreater{}\%}
  \FunctionTok{group\_by}\NormalTok{(model\_name) }\SpecialCharTok{\%\textgreater{}\%}
  \FunctionTok{nest}\NormalTok{() }

\NormalTok{fits }\OtherTok{\textless{}{-}} \FunctionTok{expand\_grid}\NormalTok{(}
\NormalTok{  analysis\_data, }
  \FunctionTok{tibble}\NormalTok{(}\AttributeTok{method =} \FunctionTok{c}\NormalTok{(}\StringTok{"AgACI"}\NormalTok{, }\StringTok{"FACI"}\NormalTok{, }\StringTok{"SF{-}OGD"}\NormalTok{, }\StringTok{"SAOCP"}\NormalTok{)), }
  \FunctionTok{tibble}\NormalTok{(}\AttributeTok{alpha =} \FunctionTok{c}\NormalTok{(}\FloatTok{0.8}\NormalTok{, }\FloatTok{0.9}\NormalTok{, }\FloatTok{0.95}\NormalTok{))}
\NormalTok{) }\SpecialCharTok{\%\textgreater{}\%}
  \FunctionTok{mutate}\NormalTok{(}\AttributeTok{fit =} \FunctionTok{pmap}\NormalTok{(}\FunctionTok{list}\NormalTok{(data, method, alpha), fit),}
         \AttributeTok{metrics =} \FunctionTok{map2}\NormalTok{(data, fit, metrics))}

\NormalTok{case\_study\_results }\OtherTok{\textless{}{-}}\NormalTok{ fits }\SpecialCharTok{\%\textgreater{}\%}
  \FunctionTok{select}\NormalTok{(}\SpecialCharTok{{-}}\NormalTok{data, }\SpecialCharTok{{-}}\NormalTok{fit) }\SpecialCharTok{\%\textgreater{}\%}
  \FunctionTok{mutate}\NormalTok{(}\AttributeTok{metrics =} \FunctionTok{map}\NormalTok{(metrics, as\_tibble)) }\SpecialCharTok{\%\textgreater{}\%}
  \FunctionTok{unnest}\NormalTok{(}\FunctionTok{c}\NormalTok{(metrics)) }
\end{Highlighting}
\end{Shaded}

\begin{figure}[H]

{\centering \includegraphics{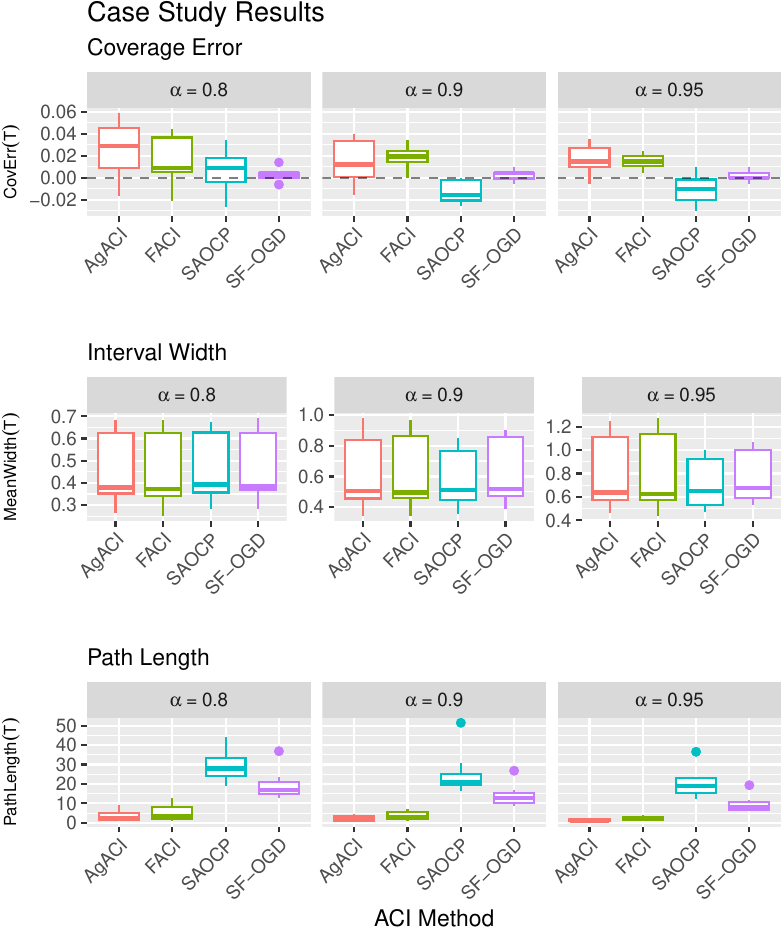}

}

\caption{\label{fig-case-study-metrics}Coverage errors, mean interval
widths, and path lengths of prediction intervals generated with each ACI
method based on forecasts from each of the 19 underlying influenza
forecasting models.}

\end{figure}

The coverage errors, mean interval widths, and path lengths of the
prediction intervals for each of the underlying forecast models is shown
in Figure~\ref{fig-case-study-metrics}. In all cases the absolute
coverage error was less than \(0.1\). SF-OGD performed particularly
well, with coverage errors close to zero for all forecasting models.
Interval widths were similar across methods, with SAOCP slightly
shorter. Path Lengths were shorter for AgACI and FACI and longer for
SAOCP.

\begin{Shaded}
\begin{Highlighting}[]
\FunctionTok{case\_study\_plot}\NormalTok{(case\_study\_results)}
\end{Highlighting}
\end{Shaded}

\begin{figure}[H]

{\centering \includegraphics{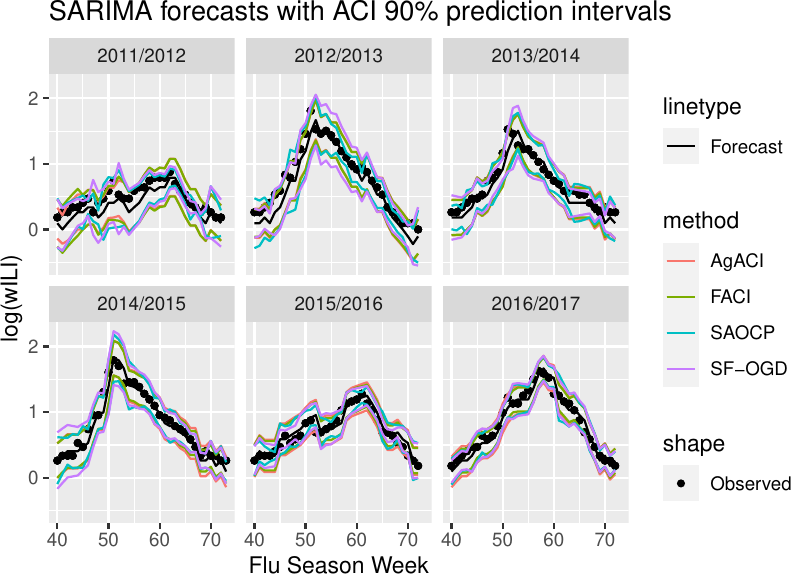}

}

\caption{\label{fig-case-study-example}Example conformal prediction
intervals for six flu seasons based on forecasts from a SARIMA type
model.}

\end{figure}

As an illustrative example, in Figure~\ref{fig-case-study-example} we
plot the point forecasts from one of the forecasting models (based on
SARIMA with no seasonal differencing) and the associated ACI-generated
90\% prediction intervals for each season from 2011-2017. In general, in
this practical setting all of the ACI algorithms yield quite similar
prediction intervals. Interestingly, the forecasts in 2011-2012
underpredicted the observations for much of the season. The algorithm
responds by making the intervals wider to cover the observations, and
because the intervals are symmetric the lower bound then becomes
unrealistically low. A similar phenomenon can be seen in the growth
phase of the 2012/2013 season as well.

\begin{Shaded}
\begin{Highlighting}[]
\NormalTok{sarima\_fits }\OtherTok{\textless{}{-}}\NormalTok{ fits }\SpecialCharTok{\%\textgreater{}\%} \FunctionTok{filter}\NormalTok{(}
\NormalTok{  model\_name }\SpecialCharTok{==} \StringTok{"ReichLab\_sarima\_seasonal\_difference\_FALSE"}\NormalTok{, }
\NormalTok{  alpha }\SpecialCharTok{==} \FloatTok{0.9}
\NormalTok{) }\SpecialCharTok{\%\textgreater{}\%}
  \FunctionTok{mutate}\NormalTok{(}\AttributeTok{output =} \FunctionTok{map}\NormalTok{(fit, extract\_intervals)) }\SpecialCharTok{\%\textgreater{}\%}
  \FunctionTok{select}\NormalTok{(method, alpha, data, output) }\SpecialCharTok{\%\textgreater{}\%}
  \FunctionTok{unnest}\NormalTok{(}\FunctionTok{c}\NormalTok{(data, output)) }\SpecialCharTok{\%\textgreater{}\%}
  \FunctionTok{filter}\NormalTok{(Season }\SpecialCharTok{!=} \StringTok{"2010/2011"}\NormalTok{)}

\NormalTok{sarima\_fits }\SpecialCharTok{\%\textgreater{}\%}
  \FunctionTok{ggplot}\NormalTok{(}\FunctionTok{aes}\NormalTok{(}\AttributeTok{x =}\NormalTok{ Model.Week, }\AttributeTok{y =} \FunctionTok{log}\NormalTok{(obs\_value))) }\SpecialCharTok{+}
  \FunctionTok{geom\_point}\NormalTok{(}\FunctionTok{aes}\NormalTok{(}\AttributeTok{shape =} \StringTok{"Observed"}\NormalTok{)) }\SpecialCharTok{+}
  \FunctionTok{geom\_line}\NormalTok{(}\FunctionTok{aes}\NormalTok{(}\AttributeTok{y =}\NormalTok{ pred, }\AttributeTok{lty =} \StringTok{"Forecast"}\NormalTok{), }\AttributeTok{color =} \StringTok{"black"}\NormalTok{) }\SpecialCharTok{+}
  \FunctionTok{geom\_line}\NormalTok{(}\FunctionTok{aes}\NormalTok{(}\AttributeTok{y =}\NormalTok{ lower, }\AttributeTok{color =}\NormalTok{ method)) }\SpecialCharTok{+}
  \FunctionTok{geom\_line}\NormalTok{(}\FunctionTok{aes}\NormalTok{(}\AttributeTok{y =}\NormalTok{ upper, }\AttributeTok{color =}\NormalTok{ method)) }\SpecialCharTok{+}
  \FunctionTok{facet\_wrap}\NormalTok{(}\SpecialCharTok{\textasciitilde{}}\NormalTok{Season) }\SpecialCharTok{+}
  \FunctionTok{labs}\NormalTok{(}
    \AttributeTok{x =} \StringTok{"Flu Season Week"}\NormalTok{, }
    \AttributeTok{y =} \StringTok{"log(wILI)"}\NormalTok{, }
    \AttributeTok{title =} \StringTok{"SARIMA forecasts with ACI 90\% prediction intervals"}
\NormalTok{  )}
\end{Highlighting}
\end{Shaded}

\hypertarget{sec-discussion}{%
\section{Discussion}\label{sec-discussion}}

The results of our simulations and case study show that, when tuning
parameters are chosen well, Adaptive Conformal Inference algorithms
yield well-performing prediction intervals. On the contrary, poor choice
of tuning parameters can lead to intervals of low utility. Furthermore,
in some cases the prediction intervals may appear to perform well with
respect to metrics like the empirical coverage error, while
simultaneously being useless in practice. The original ACI algorithm
illustrates this phenomenon: too small a value of its learning rate
\(\gamma\) yields prediction intervals that are not reactive enough,
while too large a value yields intervals that change too fast. In both
cases, the empirical coverage may appear well-calibrated, while the
prediction intervals will not be useful. Thus, the core challenge in
designing an ACI algorithm is in finding an optimal level of reactivity
for the prediction intervals. As users of these algorithms, the
challenge is in finding values for the tuning parameters that avoid
pathological behaviors.

Several of the algorithms investigated in this paper handle the problem
of finding an optimal level of reactivity by aggregating prediction
intervals generated by a set of underlying ACI algorithms. Our results
show the algorithms can perform well in multiple difficult scenarios.
However, the overall effect of these approaches is to shift the problem
to a higher level of abstraction: we still need to set tuning parameters
that control the amount of reactivity, but do so at a higher level than
the original ACI algorithm. It is desirable that these tuning parameters
be easily interpretable, with simple strategies available for setting
them. An advantage of the SF-OGD and SAOCP algorithms in this respect
are that their main tuning parameter, the maximum radius \(D\), is
easily interpretable as the maximum possible difference between the
input predictions and the truth. It is also straightforward to choose
this parameter based on a calibration set, although this strategy does
not necessarily work well in cases of distribution shift. We also found
that an advantage of the AgACI method is its robustness to the choice of
its main tuning parameter, the set of candidate learning rates. Indeed,
if AgACI does not perform well, one can simply increase the number of
candidate learning rates.

A key challenge in tuning the algorithms arises in settings of
distribution shift, where methods for choosing hyperparameters based on
a calibration set from before the distribution shift will likely not
perform well. The second simulation study we conducted probed this
setting in a simple scenario. We found that several of the methods
yielded prediction intervals that had non-optimal empirical coverage. As
we picked hyperparameters based on a calibration set formed before the
distribution shift, it is not surprising that the resulting tuning
parameters are not optimal. This underscores the difficulty in designing
ACI algorithms that can adapt to distribution shifts, and in finding
robust methods for choosing hyperparameters. In practice, it is possible
the second simulation study does not accurately reflect real-world
scenarios. Indeed, the benchmarks presented in Bhatnagar et al. (2023)
using the datasets from the M4 competition (Makridakis, Spiliotis, and
Assimakopoulos 2020), and using point predictions generated by diverse
prediction algorithms, found that ACI algorithms exhibited good
performance in terms of empirical coverage. Nevertheless, our
recommendation for future papers in this line of research is to include
simulation studies for simple distributional shift scenarios as a
benchmark.

Our case study results illustrate the dependence of the ACI algorithms
on having access to high-quality point predictions. If the predictions
are biased, for example, then the prediction intervals may be able to
achieve optimal coverage at the expense of larger interval widths. Using
ensemble methods to combine forecasts from several flexible machine
learning models is one strategy that can be used to hedge against model
misspecification and improve the quality of forecasts (Makridakis,
Spiliotis, and Assimakopoulos 2020).

There remain many possible extensions of ACI algorithms. The algorithms
presented in this work primarily consider symmetric intervals evaluated
using the pinball loss function (AgACI can yield asymmetric intervals
because the aggregation rule is applied separately to the lower and
upper bounds from the underlying experts, but those underlying experts
only produce symmetric intervals). A simple extension would switch to
using the interval loss function (Gneiting and Raftery 2007), which
would allow for asymmetric intervals where two parameters are learned
for the upper and lower bounds, respectively. It may also be of interest
to generate prediction intervals that have coverage guarantees for
arbitrary subsets of observations (for example, we may seek prediction
intervals for daily observations that have near optimal coverage for
every day of the week, or month of the year), similar to guarantees
provided by the MultiValid Prediction method described in (Bastani et
al. 2022). Another avenue for theoretical research is to propose
algorithms with provable bounds for the coverage and regret that do not
depend on the outcome being bounded.

\hypertarget{acknowledgements}{%
\subsection*{Acknowledgements}\label{acknowledgements}}
\addcontentsline{toc}{subsection}{Acknowledgements}

This research is partially supported by the Agence Nationale de la
Recherche as part of the ``Investissements d'avenir'' program (reference
ANR-19-P3IA-0001; PRAIRIE 3IA Institute). We would like to thank Margaux
Zaffran for providing helpful comments on the manuscript.

\hypertarget{references}{%
\section*{References}\label{references}}
\addcontentsline{toc}{section}{References}

\hypertarget{refs}{}
\begin{CSLReferences}{1}{0}
\leavevmode\vadjust pre{\hypertarget{ref-angelopoulos2022gentle}{}}%
Angelopoulos, Anastasios N., and Stephen Bates. 2023. {``Conformal
Prediction: A Gentle Introduction.''} \emph{Found. Trends Mach. Learn.}
16 (4): 494--591. \url{https://doi.org/10.1561/2200000101}.

\leavevmode\vadjust pre{\hypertarget{ref-bastani2022practical}{}}%
Bastani, Osbert, Varun Gupta, Christopher Jung, Georgy Noarov, Ramya
Ramalingam, and Aaron Roth. 2022. {``Practical Adversarial Multivalid
Conformal Prediction.''} In \emph{Advances in Neural Information
Processing Systems}, edited by S. Koyejo, S. Mohamed, A. Agarwal, D.
Belgrave, K. Cho, and A. Oh, 35:29362--73. Curran Associates, Inc.
\url{https://proceedings.neurips.cc/paper_files/paper/2022/file/bcdaaa1aec3ae2aa39542acefdec4e4b-Paper-Conference.pdf}.

\leavevmode\vadjust pre{\hypertarget{ref-bhatnagar2023saocp}{}}%
Bhatnagar, Aadyot, Huan Wang, Caiming Xiong, and Yu Bai. 2023.
{``Improved Online Conformal Prediction via Strongly Adaptive Online
Learning.''} In \emph{Proceedings of the 40th International Conference
on Machine Learning}. ICML'23. Honolulu, Hawaii, USA: JMLR.org.

\leavevmode\vadjust pre{\hypertarget{ref-biggerstaff2016flusight}{}}%
Biggerstaff, Matthew, David Alper, Mark Dredze, Spencer Fox, Isaac
Chun-Hai Fung, Kyle S. Hickmann, Bryan Lewis, et al. 2016. {``Results
from the Centers for Disease Control and Prevention's Predict the
2013--2014 Influenza Season Challenge.''} \emph{BMC Infectious Diseases}
16 (1): 357. \url{https://doi.org/10.1186/s12879-016-1669-x}.

\leavevmode\vadjust pre{\hypertarget{ref-cesabianchi2006games}{}}%
Cesa-Bianchi, Nicolo, and Gabor Lugosi. 2006. \emph{Prediction,
Learning, and Games}. Cambridge University Press.
\url{https://doi.org/10.1017/CBO9780511546921}.

\leavevmode\vadjust pre{\hypertarget{ref-diquigiovanni2022fd}{}}%
Diquigiovanni, Jacopo, Matteo Fontana, Aldo Solari, Simone Vantini, and
Paolo Vergottini. 2022. \emph{conformalInference.fd: Tools for Conformal
Inference for Regression in Multivariate Functional Setting}.
\url{https://CRAN.R-project.org/package=conformalInference.fd}.

\leavevmode\vadjust pre{\hypertarget{ref-feldman2023achieving}{}}%
Feldman, Shai, Liran Ringel, Stephen Bates, and Yaniv Romano. 2023.
{``Achieving Risk Control in Online Learning Settings.''}
\emph{Transactions on Machine Learning Research}.
\url{https://openreview.net/forum?id=5Y04GWvoJu}.

\leavevmode\vadjust pre{\hypertarget{ref-flusight2020}{}}%
Flusight Network. 2020. {``GitHub -
FluSightNetwork/Cdc-Flusight-Ensemble: Guidelines and Forecasts for a
Collaborative u.s. Influenza Forecasting Project.''}
\url{https://github.com/FluSightNetwork/}.

\leavevmode\vadjust pre{\hypertarget{ref-friedman1983}{}}%
Friedman, Jerome H., Eric Grosse, and Werner Stuetzle. 1983.
{``Multidimensional Additive Spline Approximation.''} \emph{SIAM Journal
on Scientific and Statistical Computing} 4 (2): 291--301.
\url{https://doi.org/10.1137/0904023}.

\leavevmode\vadjust pre{\hypertarget{ref-opera2023}{}}%
Gaillard, Pierre, Yannig Goude, Laurent Plagne, Thibaut Dubois, and
Benoit Thieurmel. 2023. \emph{Opera: Online Prediction by Expert
Aggregation}. \url{http://pierre.gaillard.me/opera.html}.

\leavevmode\vadjust pre{\hypertarget{ref-gibbs2021adaptive}{}}%
Gibbs, Isaac, and Emmanuel Candes. 2021. {``Adaptive Conformal Inference
Under Distribution Shift.''} In \emph{Advances in Neural Information
Processing Systems}, edited by M. Ranzato, A. Beygelzimer, Y. Dauphin,
P. S. Liang, and J. Wortman Vaughan, 34:1660--72. Curran Associates,
Inc.
\url{https://proceedings.neurips.cc/paper_files/paper/2021/file/0d441de75945e5acbc865406fc9a2559-Paper.pdf}.

\leavevmode\vadjust pre{\hypertarget{ref-gibbs2022faci}{}}%
Gibbs, Isaac, and Emmanuel Candès. 2022. {``Conformal Inference for
Online Prediction with Arbitrary Distribution Shifts.''}
\url{https://arxiv.org/abs/2208.08401}.

\leavevmode\vadjust pre{\hypertarget{ref-gneiting2007scoring}{}}%
Gneiting, Tilmann, and Adrian E Raftery. 2007. {``Strictly Proper
Scoring Rules, Prediction, and Estimation.''} \emph{Journal of the
American Statistical Association} 102 (477): 359--78.
\url{https://doi.org/10.1198/016214506000001437}.

\leavevmode\vadjust pre{\hypertarget{ref-gradu2022adaptive}{}}%
Gradu, Paula, Elad Hazan, and Edgar Minasyan. 2023. {``Adaptive Regret
for Control of Time-Varying Dynamics.''} In \emph{Proceedings of the 5th
Annual Learning for Dynamics and Control Conference}, edited by Nikolai
Matni, Manfred Morari, and George J. Pappas, 211:560--72. Proceedings of
Machine Learning Research. PMLR.
\url{https://proceedings.mlr.press/v211/gradu23a.html}.

\leavevmode\vadjust pre{\hypertarget{ref-krammer2018influenza}{}}%
Krammer, Florian, Gavin J. D. Smith, Ron A. M. Fouchier, Malik Peiris,
Katherine Kedzierska, Peter C. Doherty, Peter Palese, et al. 2018.
{``Influenza.''} \emph{Nature Reviews Disease Primers} 4 (1): 3.
\url{https://doi.org/10.1038/s41572-018-0002-y}.

\leavevmode\vadjust pre{\hypertarget{ref-lei2020cfcausal}{}}%
Lei, Lihua, and Emmanuel J. Candès. 2021. {``{Conformal Inference of
Counterfactuals and Individual Treatment Effects}.''} \emph{Journal of
the Royal Statistical Society Series B: Statistical Methodology} 83 (5):
911--38. \url{https://doi.org/10.1111/rssb.12445}.

\leavevmode\vadjust pre{\hypertarget{ref-lofgren2007influenza}{}}%
Lofgren, Eric, N. H. Fefferman, Y. N. Naumov, J. Gorski, and E. N.
Naumova. 2007. {``Influenza Seasonality: Underlying Causes and Modeling
Theories.''} \emph{Journal of Virology} 81 (11): 5429--36.
\url{https://doi.org/10.1128/jvi.01680-06}.

\leavevmode\vadjust pre{\hypertarget{ref-makridakis2020m4}{}}%
Makridakis, Spyros, Evangelos Spiliotis, and Vassilios Assimakopoulos.
2020. {``The M4 Competition: 100,000 Time Series and 61 Forecasting
Methods.''} \emph{International Journal of Forecasting} 36 (1): 54--74.
https://doi.org/\url{https://doi.org/10.1016/j.ijforecast.2019.04.014}.

\leavevmode\vadjust pre{\hypertarget{ref-orabona2018sfogd}{}}%
Orabona, Francesco, and Dávid Pál. 2018. {``Scale-Free Online
Learning.''} \emph{Theoretical Computer Science} 716: 50--69.
https://doi.org/\url{https://doi.org/10.1016/j.tcs.2017.11.021}.

\leavevmode\vadjust pre{\hypertarget{ref-reich2019influenza}{}}%
Reich, Nicholas G, Logan C Brooks, Spencer J Fox, Sasikiran Kandula,
Craig J McGowan, Evan Moore, Dave Osthus, et al. 2019. {``A
Collaborative Multiyear, Multimodel Assessment of Seasonal Influenza
Forecasting in the United States.''} \emph{Proc. Natl. Acad. Sci. U. S.
A.} 116 (8): 3146--54.

\leavevmode\vadjust pre{\hypertarget{ref-shafer2008conformal}{}}%
Shafer, Glenn, and Vladimir Vovk. 2008. {``A Tutorial on Conformal
Prediction.''} \emph{J. Mach. Learn. Res.} 9 (June): 371--421.

\leavevmode\vadjust pre{\hypertarget{ref-tibshirani2019ci}{}}%
Tibshirani, Ryan, Jacopo Diquigiovanni, Matteo Fontana, and Paolo
Vergottini. 2019. \emph{conformalInference: Tools for Conformal
Inference in Regression}.

\leavevmode\vadjust pre{\hypertarget{ref-tushar2019flusight}{}}%
Tushar, Abhinav, Nicholas G Reich, tkcy, brookslogan, d-osthus, Craig
McGowan, Evan Ray, et al. 2019.
{``{FluSightNetwork/cdc-flusight-ensemble: End of 2018/2019 US influenza
season}.''} Zenodo. \url{https://doi.org/10.5281/zenodo.3454212}.

\leavevmode\vadjust pre{\hypertarget{ref-tushar2018flusightnetwork}{}}%
Tushar, Abhinav, Nicholas Reich, Teresa Yamana, Dave Osthus, Craig
McGowan, Evan Ray, and et al. 2018. {``FluSightNetwork:
Cdc-Flusight-Ensemble Repository.''}
\url{https://github.com/FluSightNetwork/cdc-flusight-ensemble}.

\leavevmode\vadjust pre{\hypertarget{ref-vovk2005}{}}%
Vovk, Vladimir, Alex Gammerman, and Glenn Shafer. 2005.
\emph{Algorithmic Learning in a Random World}. Berlin, Heidelberg:
Springer-Verlag.

\leavevmode\vadjust pre{\hypertarget{ref-wintenberger2017boa}{}}%
Wintenberger, Olivier. 2017. {``Optimal Learning with Bernstein Online
Aggregation.''} \emph{Machine Learning} 106 (1): 119--41.
\url{https://doi.org/10.1007/s10994-016-5592-6}.

\leavevmode\vadjust pre{\hypertarget{ref-wright2017ranger}{}}%
Wright, Marvin N., and Andreas Ziegler. 2017. {``{ranger}: A Fast
Implementation of Random Forests for High Dimensional Data in {C++} and
{R}.''} \emph{Journal of Statistical Software} 77 (1): 1--17.
\url{https://doi.org/10.18637/jss.v077.i01}.

\leavevmode\vadjust pre{\hypertarget{ref-xu2021enbpi}{}}%
Xu, Chen, and Yao Xie. 2021. {``Conformal Prediction Interval for
Dynamic Time-Series.''} In \emph{Proceedings of the 38th International
Conference on Machine Learning}, edited by Marina Meila and Tong Zhang,
139:11559--69. Proceedings of Machine Learning Research. PMLR.
\url{https://proceedings.mlr.press/v139/xu21h.html}.

\leavevmode\vadjust pre{\hypertarget{ref-xu2023spci}{}}%
---------. 2023. {``Sequential Predictive Conformal Inference for Time
Series.''} In \emph{Proceedings of the 40th International Conference on
Machine Learning}, edited by Andreas Krause, Emma Brunskill, Kyunghyun
Cho, Barbara Engelhardt, Sivan Sabato, and Jonathan Scarlett,
202:38707--27. Proceedings of Machine Learning Research. PMLR.
\url{https://proceedings.mlr.press/v202/xu23r.html}.

\leavevmode\vadjust pre{\hypertarget{ref-zaffran2022agaci}{}}%
Zaffran, Margaux, Olivier Feron, Yannig Goude, Julie Josse, and Aymeric
Dieuleveut. 2022. {``Adaptive Conformal Predictions for Time Series.''}
In \emph{Proceedings of the 39th International Conference on Machine
Learning}, edited by Kamalika Chaudhuri, Stefanie Jegelka, Le Song,
Csaba Szepesvari, Gang Niu, and Sivan Sabato, 162:25834--66. Proceedings
of Machine Learning Research. PMLR.
\url{https://proceedings.mlr.press/v162/zaffran22a.html}.

\end{CSLReferences}

\hypertarget{appendix}{%
\section{Appendix}\label{appendix}}

\hypertarget{additional-simulation-study-results}{%
\subsection{Additional simulation study
results}\label{additional-simulation-study-results}}

\begin{Shaded}
\begin{Highlighting}[]
\FunctionTok{simulation\_one\_plot}\NormalTok{(simulation\_study1}\SpecialCharTok{$}\NormalTok{results)}
\end{Highlighting}
\end{Shaded}

\begin{figure}[H]

{\centering \includegraphics{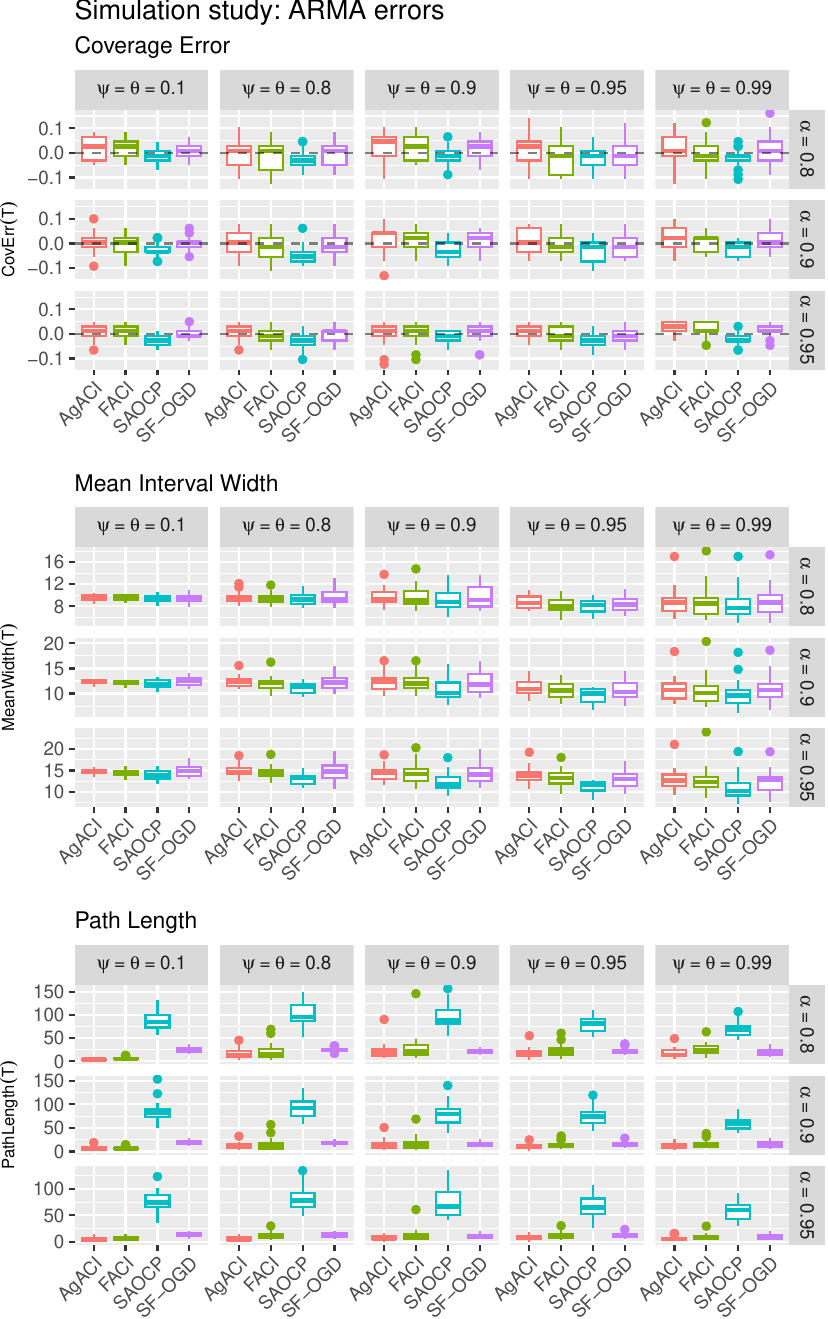}

}

\caption{Coverage errors, mean interval widths, and path lengths for the
first simulation study with target coverage
\(\alpha \in \{ 0.8, 0.9, 0.95 \}\).}

\end{figure}

\begin{Shaded}
\begin{Highlighting}[]
\FunctionTok{simulation\_one\_joint\_plot}\NormalTok{(simulation\_study1}\SpecialCharTok{$}\NormalTok{results)}
\end{Highlighting}
\end{Shaded}

\begin{figure}[H]

{\centering \includegraphics{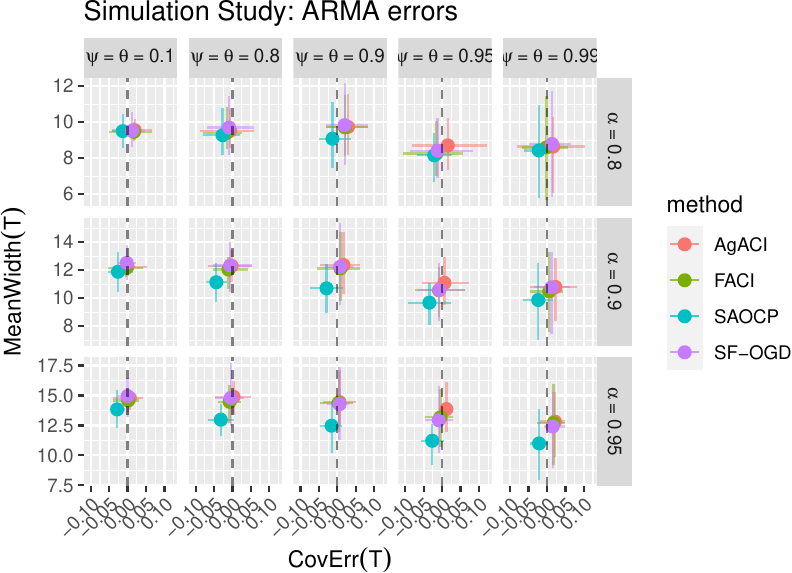}

}

\caption{\label{fig-simulation-one-joint}Mean Interval Width vs Coverage
Error for the first simulation study. The error bars represent the 10\%
to 90\% quantiles of the metrics over the simulation datasets.}

\end{figure}

\begin{Shaded}
\begin{Highlighting}[]
\FunctionTok{simulation\_two\_plot}\NormalTok{(simulation\_study2}\SpecialCharTok{$}\NormalTok{results)}
\end{Highlighting}
\end{Shaded}

\begin{figure}[H]

{\centering \includegraphics{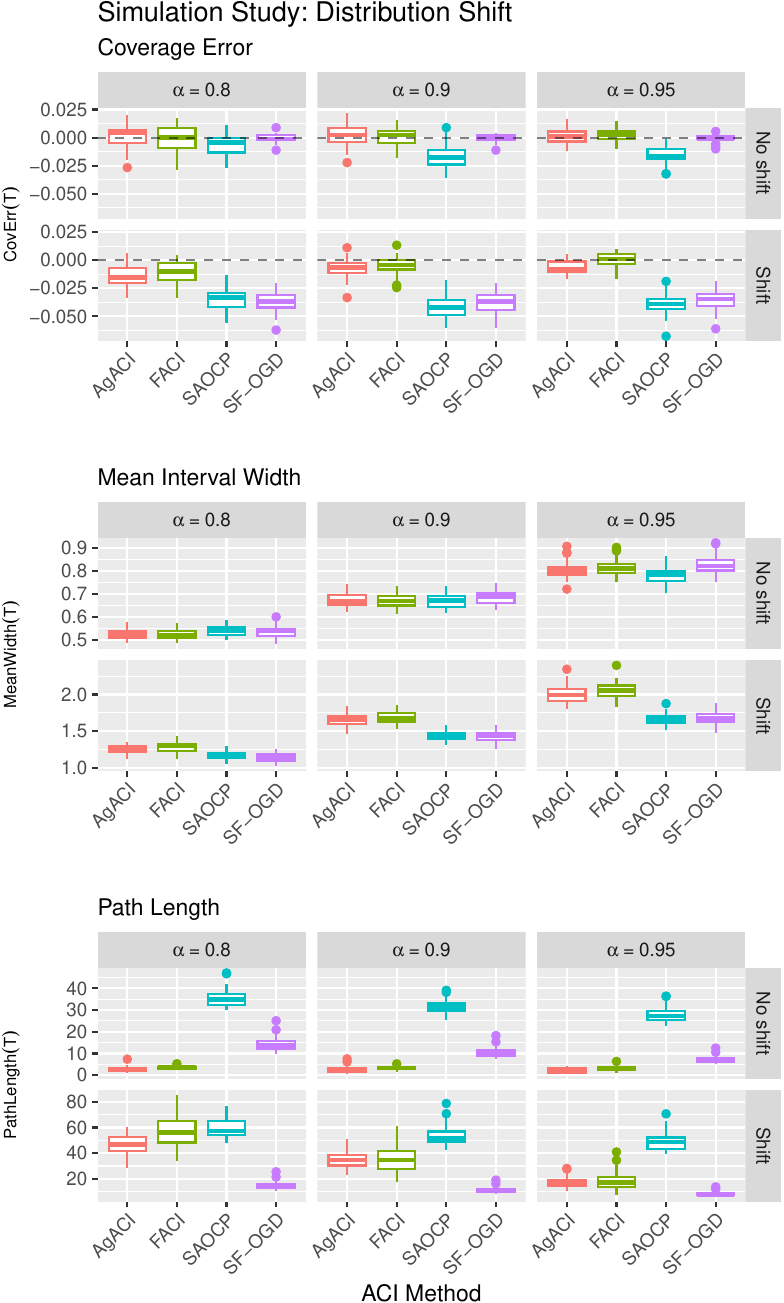}

}

\caption{\label{fig-simulation-two-results}Coverage error, mean interval
width, and path length for \(\alpha = 0.8, 0.9, 0.95\) and simulations
with and without distributional shift.}

\end{figure}

\end{document}